\documentclass[journal]{IEEEtran}
\usepackage{cite}
\ifCLASSINFOpdf
  \usepackage[pdftex]{graphicx}
  \graphicspath{{./pdf/}{./jpeg/}}
  \DeclareGraphicsExtensions{.pdf,.jpeg,.png}
\else
  \usepackage[dvips]{graphicx}
  \graphicspath{{./eps/}}
  \DeclareGraphicsExtensions{.eps}
\fi
\usepackage{amsmath}
\usepackage{algorithmic}
\usepackage{array}
\ifCLASSOPTIONcompsoc
  \usepackage[caption=false,font=normalsize,labelfont=sf,textfont=sf]{subfig}
\else
  \usepackage[caption=false,font=footnotesize]{subfig}
\fi
 \usepackage{dblfloatfix}
\usepackage{url}
\usepackage{amssymb,amsthm,dsfont}
\usepackage{enumerate}
\usepackage[hidelinks]{hyperref}

\def\markov{\mathrel{\mathpalette\dash\circ}}
\def\dash#1#2{\ooalign{$#1-\mkern-10mu-$\crcr$\hfil#1#2\hfil$}}

\def\ind{\mathds{1}}

\def\EE{\mathbb{E}}
\def\NN{\mathbb{N}}

\def\RR{\mathbb{R}}
\def\ZZ{\mathbb{Z}}

\def\e{{\rm e}}

\def\iid{{\rm i.i.d.}}
\def\cc{{\rm c.c.}}

\def\dotle{\mathrel{\dot{\le}}}

\def\dotge{\mathrel{\dot{\ge}}}

\def\ale{\stackrel{\text{(a)}}{\le}}
\def\ble{\stackrel{\text{(b)}}{\le}}
\def\cle{\stackrel{\text{(c)}}{\le}}

\def\adotle{\stackrel{\text{(a)}}{\dotle}}

\def\age{\stackrel{\text{(a)}}{\ge}}
\def\bge{\stackrel{\text{(b)}}{\ge}}
\def\cge{\stackrel{\text{(c)}}{\ge}}

\def\aeq{\stackrel{\text{(a)}}{=}}
\def\beq{\stackrel{\text{(b)}}{=}}

\def\ddoteq{\stackrel{\text{(d)}}{\doteq}}

\def\bdotge{\stackrel{\text{(b)}}{\dotge}}

\DeclareMathOperator{\E}{\EE}
\DeclareMathOperator{\var}{var}
\DeclareMathOperator{\cov}{cov}

\DeclareMathOperator*{\argmin}{arg\,min}
\DeclareMathOperator*{\supp}{supp}

\def\cA{{\cal A}}
\def\cC{{\cal C}}
\def\cE{{\cal E}}
\def\cK{{\cal K}}
\def\cP{{\cal P}}
\def\cQ{{\cal Q}}
\def\cS{{\cal S}}
\def\cT{{\cal T}}
\def\cU{{\cal U}}
\def\cV{{\cal V}}

\def\cX{{\cal X}}
\def\cY{{\cal Y}}
\def\cZ{{\cal Z}}
\def\rE{{\rm E}}
\def\rM{{\rm M}}
\def\rd{{\rm d}}
\def\re{{\rm e}}
\def\rr{{\rm r}}
\def\rs{{\rm s}}

\def\abs#1{{\lvert#1\rvert}}
\def\bigabs#1{{\bigl\lvert#1\bigr\rvert}}

\def\type#1{\hat{Q}_{{#1}}}

\newtheorem{theorem}{Theorem}
\newtheorem{corollary}[theorem]{Corollary}
\newtheorem{lemma}[theorem]{Lemma}
\theoremstyle{definition}

\newtheorem{definition}{Definition}
\theoremstyle{remark}
\newtheorem{remark}{Remark}
\newtheorem*{remark*}{Remark}

\begin{document}
%
% paper title
% Titles are generally capitalized except for words such as a, an, and, as,
% at, but, by, for, in, nor, of, on, or, the, to and up, which are usually
% not capitalized unless they are the first or last word of the title.
% Linebreaks \\ can be used within to get better formatting as desired.
% Do not put math or special symbols in the title.
\title{Exact Random Coding Secrecy Exponents for the Wiretap Channel}
%
%
% author names and IEEE memberships
% note positions of commas and nonbreaking spaces ( ~ ) LaTeX will not break
% a structure at a ~ so this keeps an author's name from being broken across
% two lines.
% use \thanks{} to gain access to the first footnote area
% a separate \thanks must be used for each paragraph as LaTeX2e's \thanks
% was not built to handle multiple paragraphs
%

\urldef{\mani}\url{mani.bastaniparizi@epfl.ch}
\urldef{\emre}\url{emre.telatar@epfl.ch}
\urldef{\neri}\url{merhav@ee.technion.ac.il}

\author{Mani~Bastani~Parizi,~\IEEEmembership{Student~Member,~IEEE,}
  Emre~Telatar,~\IEEEmembership{Fellow,~IEEE,}
  and~Neri~Merhav,~\IEEEmembership{Fellow,~IEEE}% <-this % stops a space
  \thanks{The authors would like to thank anonymous reviewers for their
  helpful comments that improved the quality of the manuscript.}%
  \thanks{The work of M.~Bastani~Parizi and E.~Telatar was supported by the
    Swiss National Science Foundation (SNSF) grant no.~200020\_146832.  The
    work of N.~Merhav was supported by the Israel Science Foundation (ISF),
  grant no.~412/12.}%
  \thanks{The material in this paper was presented in part in 2016 IEEE
  International Symposium on Information Theory (ISIT 2016).}%
  \thanks{M.~Bastani~Parizi and E.~Telatar are with the Information Theory
    Laboratory (LTHI), Swiss Federal Institute of Technology (EPFL),
    Lausanne 1015, Switzerland (email: \mani, \emre)}% <-this % stops a space 
  \thanks{N.~Merhav is with the Department of Electrical Engineering,
    Technion - Israel Institute of Technology, Haifa 32000, Israel (email:
  \neri)}% <-this % stops a space
}

\maketitle

% As a general rule, do not put math, special symbols or citations
% in the abstract or keywords.
\begin{abstract}
  We analyze the exact exponential decay rate of the expected amount of
  information leaked to the wiretapper in Wyner's wiretap channel setting
  using wiretap channel codes constructed from both i.i.d.\ and
  constant-composition random codes.  Our analysis for those  sampled from
  i.i.d.\ random coding ensemble shows that the previously-known achievable
  secrecy exponent using this ensemble is indeed the exact exponent for an
  average code in the ensemble.  Furthermore, our analysis on wiretap
  channel codes constructed from the ensemble of constant-composition
  random codes leads to an exponent which, in addition to being the exact
  exponent for an average code, is larger than the achievable secrecy
  exponent that has been established so far in the literature for this
  ensemble (which in turn was known to be smaller than that achievable by
  wiretap channel codes sampled from i.i.d.\ random coding ensemble).  We
  show examples where the exact secrecy exponent for the wiretap
  channel codes constructed from random constant-composition codes is
  larger than that of those constructed from i.i.d.\ random codes and
  examples where the exact secrecy exponent for the wiretap channel codes
  constructed from i.i.d.\ random codes is larger than that of those
  constructed from constant-composition random codes.  We, hence, conclude
  that, unlike the error correction problem, there is no general ordering
  between the two random coding ensembles in terms of their secrecy
  exponent.
\end{abstract}

% Note that keywords are not normally used for peerreview papers.
\begin{IEEEkeywords}
  Wiretap channel, Channel resolvability, Secrecy exponent, Resolvability
  exponent
\end{IEEEkeywords}

% For peer review papers, you can put extra information on the cover
% page as needed:
% \ifCLASSOPTIONpeerreview
% \begin{center} \bfseries EDICS Category: 3-BBND \end{center}
% \fi
%
% For peerreview papers, this IEEEtran command inserts a page break and
% creates the second title. It will be ignored for other modes.
\IEEEpeerreviewmaketitle

\section{Introduction} 
\begin{figure*}[htb]
  \centerline{\includegraphics{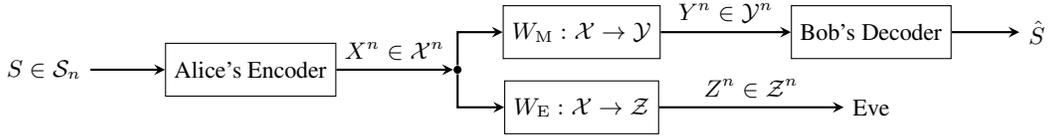}}
  \caption{Wiretap Channel}
  \label{fig:wiretap}
\end{figure*}

\IEEEPARstart{T}{he} problem of communication in presence of an
eavesdropper wiretapping the signals sent to the legitimate receiver (see
Figure~\ref{fig:wiretap}) was first studied by Wyner \cite{wyner:1975a} and
later, in a broader context, by Csisz\'ar and K\"orner \cite{csiszar:1978},
where it was shown (among other results) that as long as the eavesdropper's
channel is weaker than legitimate receiver's channel, reliable and
\emph{secure} communication at positive rates is feasible.  More precisely,
it was shown that, given any distribution on the common input alphabet of
the channels, $P_X$, for which the mutual information developed across the
legitimate receiver's channel is higher than that developed across the
wiretapper's channel, that is, $I(X;Y)>I(X;Z)$, with $(X,Y,Z)\sim
P_X(x)W_\rM(y|x)W_\rE(z|x)$ (where $X$, $Y$, and $Z$ represent the common
  input, legitimate receiver's channel output, and wiretapper's channel
output, respectively), as long as the secret message rate $R_\rs \triangleq
\frac1n \log\abs{\cS_n}$ is below $I(X;Y)-I(X;Z)$ there exists a sequence of
coding schemes (indexed by the block-length $n$) using which
\begin{subequations} 
  \begin{align} \lim_{n\to\infty} 
    \max_{s \in \cS_n} \Pr\{\hat{s}_{\rm ML}(Y^n) \ne S |S=s\}&=0, \\ 
    \lim_{n\to\infty} \frac1n I(S;Z^n) &=0.  \label{eq:weaksec}
  \end{align} 
\end{subequations}
In the above, $S$ represents the secret message taking values in the 
message set $\cS_n$, $\hat{s}_{\rm ML}(Y^n)$ is the maximum-likelihood (ML)
estimation of the sent message given the output sequence of the legitimate
receiver's channel and $Z^n$ represents the
output sequence of the wiretapper's channel (see Figure~\ref{fig:wiretap}).

Classical codes for the wiretap channel are constructed by associating each
message with a code that operates at a rate $R$ \emph{just below}
the mutual information developed across the eavesdropper's channel.  To
communicate a message, the stochastic encoder of Alice picks a codeword
uniformly at random from the code associated to that message and transmits
it via consecutive uses of the channel
\cite{wyner:1975a,csiszar:1978,massey:1983}.  Such constructions, known as
\emph{capacity-based constructions} (with a slight abuse of terminology)
\cite{bloch:2013}, will guarantee that the normalized amount of information
that Eve learns about the secret message by observing her channel output
signal, $\frac1nI(S;Z^n)$, will be arbitrarily small, provided that the
block-length $n$ is sufficiently large.  Recently,
\emph{resolvability-based} constructions for wiretap channel codes, namely,
those associating each message with a code operating at a rate
\emph{just above} the mutual information of the wiretapper's channel was
shown to be more powerful than the capacity-based constructions to prove
achievability results. Indeed, in \cite{csiszar:1996} it was shown that
such constructions can be used to easily show that the \emph{unnormalized}
amount of information Eve learns about the secret message, $I(S;Z^n)$,
vanishes as the block-length increases, namely to
establish \emph{strong secrecy} (a notion first introduced by Maurer and
Wolf \cite{maurer:2000}). In particular, when resolvability-based wiretap
channel codes are employed over stationary memoryless wiretap channels the
amount of information Eve learns about the secret message vanishes
\emph{exponentially fast} in the block-length.  Thus, it is natural to
study the rate of this exponential decay.

\begin{definition}\label{def:secexp}
  Given the rate pair $(R_\rs,R)$ and a pair of stationary memoryless
  channels $(W_\rM,W_\rE)$, a number $\eta$ is an achievable
  \emph{secrecy exponent} if there
  exists a sequence of coding schemes of block-length $n$ and secret
  message rate $R_\rs$, each message associated with a sub-code of rate $R$
  (i.e., the encoder needs access to a random number generator of rate $R$)
  that are reliable for communication over $W_\rM$ and guarantee
  \begin{equation} 
    \liminf_{n\to\infty} -\frac1n \log I(S;Z^n) \ge \eta.
  \end{equation}
\end{definition}

Hayashi \cite{hayashi:2006} was the first to derive a lower bound to the
achievable secrecy exponents using the resolvability-based construction of
wiretap channel codes from i.i.d.\ random codes. He, later on, showed that
this lower bound can be improved if, on top of a random code sampled from
i.i.d.\ random coding ensemble, a random hash function is used in the
construction of the encoder--decoder pair \cite{hayashi:2011a}. This
technique is known as \emph{privacy amplification}. More recently, it was
shown (see special cases of \cite[Theorem~2]{hayashi:2012},
\cite[Theorem~3.1]{han:2014}, or the proof given in \cite{bastani:2015})
that privacy amplification is unnecessary and the exponent derived in
\cite{hayashi:2011a} lower-bounds the exponential decay rate of the
ensemble average of the information leaked to Eve when a wiretap channel
code constructed from the ensemble of i.i.d.\ random codes is used for
communication.

To study the \emph{universally achievable} (in the sense defined in
\cite{korner:1980}) secrecy exponents, constructing codes for wiretap
channel from the ensemble of random constant-composition codes is
investigated in \cite{hayashi:2011b}.  A lower bound to the
achievable secrecy exponent when this class of wiretap channel codes are
used in conjunction with privacy amplification is derived in
\cite{hayashi:2011b} which is smaller than the lower bound of
\cite{hayashi:2011a} on the achievable secrecy exponent using i.i.d.\
random codes.
%%%%%%%%%%%%%%%%%%%%%%%%%%%%%%%%%%%%%%%%%%%%%%%%%%%%%%%%%%%%%%%%%%%%%%%%%%%
\subsection{Contribution and Paper Outline}
In this paper we first show that the exponent derived via the method of
\cite{bastani:2015} (which was first established in \cite{hayashi:2011a})
is indeed the \emph{exact} secrecy exponent for an average code in the
ensemble and secondly extend the analysis of \cite{bastani:2015} to the
ensemble of constant-composition random codes (see
Theorem~\ref{thm:asymptotic} and its corollary).  This, in particular,
implies that the previously-known lower bound to the achievable secrecy
exponent using wiretap channel codes constructed from i.i.d.\ random coding
ensemble characterizes the exact exponential decay rate of the average
amount of information leaked to the eavesdropper. Moreover, it turns out
that the exact secrecy exponent for the wiretap channel codes constructed
from constant-composition random codes is larger than the lower bound
derived in \cite{hayashi:2011b} and there are examples where this dominance
is strict. Further, examples show that in general there is no ordering
between the secrecy exponents of the ensembles of i.i.d.\ and
constant-composition codes.  In other words, for some channels the i.i.d.\
ensemble yields a better secrecy exponent, whereas in the others, the
constant-composition ensemble prevails (see Section~\ref{sec:compare}).

The analysis of \cite{bastani:2015} is based on pure random coding
arguments (no privacy amplification is used) and is carried out by
lower-bounding the achievable \emph{resolvability exponents} (see
Definition~\ref{def:resexp}) using random codes.  We will show, in this
work, that this method not only proves the achievability of the exponent,
but also, using very similar steps, establishes its exactness (see
Definition~\ref{def:exactResExp}).  Moreover, a simple observation
shows that the exact resolvability exponent equals the exact secrecy
exponent for an ensemble (see Theorem~\ref{thm:exactSecExp}), which in turn,
allows us to conclude that the exponent derived through this method is the
exact secrecy exponent as well.

The remainder of this paper is organized as follows.  After setting our
notation conventions in Section~\ref{sec:notation}, we prove the
equivalence of secrecy and resolvability exponents in Section~\ref{sec:res}
and reduce the analysis of the exact secrecy exponent for an ensemble to
that of the exact resolvability exponent.  We present our main result on
exact secrecy exponents in Section~\ref{sec:exp}, argue that the exact
secrecy exponent for the ensemble of constant-composition random codes is
larger than the lower bound derived in \cite{hayashi:2011b}, and give
numerical examples comparing the exponents for two ensembles of i.i.d.\ and
constant-composition random codes.  Our main result is proved in
Section~\ref{sec:proof}.  To streamline the presentation, we relegate the
straightforward but tedious parts of the proof to the appendices.
%%%%%%%%%%%%%%%%%%%%%%%%%%%%%%%%%%%%%%%%%%%%%%%%%%%%%%%%%%%%%%%%%%%%%%%%%%%
\subsection{Related Work}
In addition to those cited above, \cite{hou:2013} also presents a simple
achievability proof for channel resolvability.  Based on this proof the
authors, in their subsequent work \cite{hou:2014}, establish strong secrecy
for wiretap channel using resolvability-based constructions for wiretap
channel codes.  The performance of a code for the wiretap channel is
measured via two figures of merit, namely, the error probability and
information leakage, both of which decay exponentially in block-length when
a wiretap channel code sampled from the ensemble of random codes is
employed on stationary memoryless channels (as we will also discuss
in Theorem~\ref{thm:existence}). The trade-off between secrecy and error
exponents (as well as other generalizations of the model) is studied in
\cite{chou:2015}.

Another important problem, in the realm of information-theoretic secrecy, is
\emph{secret key agreement} \cite{maurer:1993,ahlswede:1993}.  The secrecy
exponents related to this model are studied in
\cite{hayashi:2011a,hayashi:2013,chou:2015,hayashi:2015} and, in
particular, in \cite{hayashi:2013,hayashi:2015} shown to be exact.
%%%%%%%%%%%%%%%%%%%%%%%%%%%%%%%%%%%%%%%%%%%%%%%%%%%%%%%%%%%%%%%%%%%%%%%%%%%
\section{Notation}\label{sec:notation}
We use uppercase letters (like $X$) to denote a random variable and
the corresponding lowercase version ($x$) for a realization of that random
variable.  The same convention applies to vectors, i.e., $x^n =
(x_1,\dotsc,x_n)$ denotes a realization of the random vector
$X^n=(X_1,\dotsc,X_n)$.  We denote finite sets by script-style uppercase
letters like $\cA$. The cardinality of the set $\cA$ is denoted by
$\abs{\cA}$.

We write $f(n) \dotle g(n)$ if there exists a function $p(n)$ such that
$\limsup_{n\to\infty} \frac1n \log(p(n)) = 0$ and $f(n) \le
p(n) g(n)$.  As noted in \cite[p.~2507]{csiszar:1998}, when $f(n)$ and
$g(n)$ depend on other variables than $n$ it is understood that $p(n)$ can
only depend on the \emph{fixed parameters} of the problem such as channel
transition probabilities, the cardinality of its input and output alphabet,
and its input distribution and not the other parameters $f$ and $g$ may
depend on.\footnote{Let
$\theta$ be a parameter that $f$ and $g$ depend on.  If $f_\theta(n) \dotle
g_\theta(n)$ then, $\forall\theta$, $\limsup_{n\to\infty} \frac1n
\log\left( \frac{f_\theta(n)}{g_\theta(n)}\right) \le 0$ but the reverse is
not true.  In fact $f_\theta(n) \dotle g_\theta(n)$ is equivalent to
$\limsup_{n\to\infty} \sup_{\theta} \frac1n
\log\left(\frac{f_\theta(n)}{g_\theta(n)}\right) \le 0$ which is a stronger
statement than the former.} $f(n) \doteq g(n)$ means $f(n) \dotle g(n)$ and
$g(n) \dotle f(n)$.  
For $a \in \RR$, $[a]^+ \triangleq \max\{a, 0\}$ denotes positive clipping.

We denote the set of distributions on alphabet $\cX$ as $\cP(\cX)$. If
$P\in\cP(\cX)$,  $P^n\in\cP(\cX^n)$ denotes the product distribution
$P^n(x^n)\triangleq\prod_{i=1}^nP(x_i)$ (where $x^n$ denotes the
$n$-dimensional vector $(x_1,\dotsc,x_n)\in\cX^n$).  Likewise, if
$V\colon\cX\to\cY$ is a conditional distribution (that is, $\forall
x\in\cX$, $V(\cdot|x)\in\cP(\cY)$), $V^n\colon\cX^n\to\cY^n$ denotes the
conditional distribution $V^n(y^n|x^n)=\prod_{i=1}^nV(y_i|x_i)$.  For a
joint distribution $Q\in\cP(\cX\times\cY)$, $Q_X$ (respectively $Q_Y$)
denotes its $x$- (respectively  $y$-) marginal.  For $P\in\cP(\cX)$ and a
stochastic matrix $V\colon\cX\to\cY$, $P\times V\in\cP(\cX\times\cY)$
denotes the joint distribution $P(x)V(y|x)$ and $P\circ V \in \cP(\cY)$
denotes the $y$-marginal of the joint distribution $P\times V$, that is
$(\!P\circ V\!)(y)=(\!P\times V\!)_Y(y)=\sum_{x}P(x)V(y|x)$.

We denote the \emph{type} of a sequence $x^n \in \cX^n$ by
$\type{x^n}\in\cP(\cX)$.  A distribution $P \in \cP(\cX)$ is an
\emph{$n$-type} if $\forall x\in\cX\colon nP(x)\in\ZZ$.  We denote the
set of $n$-types on $\cX$ as ${\cP_n}(\cX)\subsetneq\cP(\cX)$ and use the
fact that $\abs{\cP_n(\cX)}\le(n+1)^\abs{\cX}$ \cite[Lemma~2.2]{csiszar:it}
repeatedly.  If $P\in{\cP}_n(\cX)$, we denote the set of all
sequences of type $P$ as $\cT^n_P\subset\cX^n$. 

For a distribution $P\in\cP(\cX)$, $\supp(P)\triangleq\{x\in\cX\colon P(x)
> 0\}$. If $P,Q\in\cP(\cX)$ are a pair of distributions we say $P$ is absolutely
continuous with respect to $Q$, and denote this by $P\ll Q$, if
$\supp(P)\subseteq\supp(Q)$.

The $\ell_1$ distance and divergence between two distributions
$P,Q\in\cP(\cX)$ are, respectively, defined as
\begin{equation}
  \abs{P-Q}\triangleq\sum_{x\in\cX}\abs{P(x)-Q(x)}
\end{equation}
and
\begin{equation}
  D(P\|Q)\triangleq\sum_{x\in\cX}P(x)\log\frac{P(x)}{Q(x)}
\end{equation}
(here and in the sequel the bases of $\log$ and $\exp$ are arbitrary but
the same).  For two stochastic matrices $V\colon\cX\to\cY$ and
$W\colon\cX\to\cY$, and $P\in\cP(\cX)$, the conditional divergence is
defined as
\begin{align}
  D(V\|W|P) 
  &\triangleq\sum_{x\in\cX}P(x)\sum_{y\in\cY}V(y|x)\log\frac{V(y|x)}{W(y|x)}
  \\
  & = D(P\times V\|P \times W).
\end{align}
For $P\in\cP(\cX)$, 
\begin{equation}
  H(P)\triangleq-\sum_{x\in\cX}P(x)\log P(x).
\end{equation}
For $Q\in\cP(\cX\times\cY)$, $I(Q) \triangleq D(Q\|Q_X\times Q_Y)$.  If
$P\in\cP(\cX)$ and $V\colon\cX\to\cY$ is a stochastic matrix,
$I(P,V)\triangleq I(P\times V)$ denotes the mutual information developed
across the channel $V$ with input distribution $P$.
%%%%%%%%%%%%%%%%%%%%%%%%%%%%%%%%%%%%%%%%%%%%%%%%%%%%%%%%%%%%%%%%%%%%%%%%%%%
\section{Secrecy via Channel Resolvability} \label{sec:res} 
As we mentioned earlier, \emph{channel resolvability} is a convenient and
powerful tool for the analysis of secrecy \cite{csiszar:1996,bloch:2013}.
The concept of resolvability dates back to Wyner \cite{wyner:1975b}, where
he observed that, given a stationary memoryless channel $W\colon\cX\to\cZ$
and an input distribution $P_X$ that induces the distribution $P_Z=P_X\circ
W$ at its output, it is possible to well-approximate the product
distribution $P_Z^n$ at the output of $W^n$ (the product channel
corresponding to $n$ independent uses of $W$) by transmitting a uniformly
chosen codeword from a code of rate $R > I(X;Z)$.  Indeed, if the code is
sampled from the i.i.d.\ random coding ensemble, with very high probability
the normalized divergence between the channel output distribution and
$P_Z^n$ can be made arbitrarily small by choosing $n$ sufficiently large.
Han and Verd\'u \cite{han:1993} and Hayashi \cite{hayashi:2006} developed
this theory further by replacing the measure of approximation by normalized
$\ell_1$ distance and unnormalized divergence, respectively, and showed
first, that the same limits on the code size hold in these cases and,
second, that the distance between the output distribution and the target
distribution $P_Z^n$ vanishes exponentially fast as the block-length
increases (similar results are derived in
\cite{cuff:2013,hou:2013,bastani:2015} as well).  In particular, in
\cite{hayashi:2006,hou:2014,han:2014,bastani:2015}, the exponential decay
of the informational divergence is leveraged to establish an exponentially
decaying upper bound on the information leaked to the eavesdropper in
wiretap channel's model.  

We can extend the notion of resolvability and ask for the approximation of
arbitrary target distributions.  Given a code $\cC_n=\{x^n_1,\dotsc,x^n_M\}$
(of block-length $n$ and size $M$) and the channel $W\colon\cX\to\cZ$,
denote by $P_{\cC_n}$ the output distribution of $W^n$ when a uniformly
chosen codeword from $\cC_n$ is transmitted, that is,
\begin{equation} \label{eq:output}
  P_{\cC_n}(z^n) \triangleq \frac1M \sum_{i=1}^{M} W^n(z^n|x^n_i).
\end{equation}

\begin{definition}
  Given a stationary memoryless channel $W\colon\cX\to\cZ$, a rate $R$, and
  a sequence of target distributions $\Phi=\{\Phi_n \in
  \cP(\cZ^n)\}_{n\in\NN}$, a number $E^{\Phi}(W,R)$ is an achievable
  \emph{resolvability exponent} over the channel $W$, at rate $R$, with
  respect to $\Phi$ if there exists a sequence $\{\cC_n\}_{n\in\NN}$ of
  codes ($\cC_n$ of block-length $n$), such that
  $\limsup_{n\to\infty}\frac1n\log\abs{\cC_n}\le R$ and 
  \begin{equation}
    \liminf_{n\to\infty} -\frac1n \log D(P_{\cC_n}\|\Phi_n) \ge
    E^{\Phi}(W,R).
  \end{equation}
\end{definition}
\begin{definition}
  The supremum of all achievable resolvability exponents over
  $W\colon\cX\to\cZ$, at rate $R$, with respect to
  $\Phi=\{\Phi_n\in\cP(\cZ^n)\}_{n\in\NN}$ is \emph{the}
  resolvability exponent of the channel $W\colon\cX\to\cZ$ at rate $R$ with
  respect to $\Phi$.  
\end{definition}

Computing ``the'' resolvability exponent is a difficult task as it
necessitates a search over all possible sequences of codes to find the
best resolvability code.  The usual way to circumvent such a difficulty is
to use the probabilistic method and analyze the achievable exponents for
an ensemble of random codes. 
\begin{definition}\label{def:randomcode}
  Given $\Pi=\{P_{X^n}\in\cP(\cX^n)\}_{n\in\NN}$, a sequence of probability
  distributions on $\cX^n$, an \emph{ensemble of random codes} of rate (at
  most) $R$ is a sequence of random codes $\cC_n$ of block-length $n$ and
  size $M=\lfloor\exp(nR)\rfloor$ obtained by sampling the codewords
  independently from the distribution $P_{X^n}$.  In other words, 
  \begin{equation} \label{eq:randomcode}
    \Pr\bigl\{\cC_n = \{x^n_1,\dotsc,x^n_M\}\bigr\} = 
    \prod_{i=1}^{M} P_{X^n}(x^n_i).
  \end{equation}
\end{definition}
\begin{definition} \label{def:resexp}
  Given $\Pi=\{P_{X^n}\in\cP(\cX^n)\}_{n\in\NN}$, a stationary memoryless
  channel $W\colon\cX\to\cZ$, and a rate $R$, a number
  $\underline{E}_\rs(\Pi,W,R)$ is an achievable resolvability exponent for
  the ensemble of random codes of rate (at most) $R$ defined by $\Pi$, over
  the channel $W$, if 
  \begin{equation}
    \liminf_{n\to\infty} -\frac1n \log \E[D(P_{\cC_n}\|\bar{P}_{Z^n})] \ge
    \underline{E}_\rs(\Pi,W,R),
  \end{equation}
  where $\cC_n$ is a random code of size $M=\lfloor\exp(nR)\rfloor$
  distributed according to \eqref{eq:randomcode} and the sequence of target
  distributions $\{\bar{P}_{Z^n}\in\cP(\cZ^n)\}_{n\in\NN}$ is defined as
  \begin{equation}
    \bar{P}_{Z^n}(z^n) \triangleq (\!P_{X^n} \circ W^n\!)(z^n) = \sum_{x^n\in\cX^n}
    P_{X^n}(x^n) W^n(z^n|x^n).
    \label{eq:ref}
  \end{equation}
\end{definition}
\begin{remark*} 
  In the passage to the probabilistic method, we restricted the sequence of
  target measures to those induced by the code sampling
  distribution $P_{X^n}$ at the output of the $n$-fold use of $W$,
  \eqref{eq:ref}.
  Indeed, it is easy to verify that when $\cC_n$ is a random code whose
  codewords are drawn independently from $P_{X^n}$, for any distribution
  $\Phi_n\in\cP(\cZ^n)$, 
  \begin{equation}
    \E[D(P_{\cC_n}\|\Phi_n)] = \E[D(P_{\cC_n}\|\bar{P}_{Z^n})] +
    D(\bar{P}_{Z^n}\|\Phi_n).
  \end{equation}
  Therefore, to show the existence of good resolvability
  codes for approximating a sequence of target distributions
  $\{\Phi_n\in\cP(\cZ^n)\}_{n\in\NN}$ via random coding arguments, we can
  exclusively consider the ensembles of random codes whose sampling
  distribution $P_{X^n}$ induces $\Phi_n$ at the output of $W^n$---any
  other ensemble is \emph{suboptimal} due to the residual divergence
  $D(\bar{P}_{Z^n}\|\Phi_n)$.
\end{remark*}
\begin{definition}\label{def:exactResExp}
  The \emph{exact} resolvability exponent of the ensemble of random codes
  of rate (at most) $R$ defined via the sequence of distributions
  $\Pi=\{P_{X^n}\in\cP(\cX^n)\}_{n\in\NN}$, over the channel
  $W\colon\cX\to\cZ$, is defined as
  \begin{equation}
    E_\rs(\Pi,W,R) \triangleq \lim_{n\to\infty} -\frac1n \log
    \EE[D(P_{\cC_n}\|\bar{P}_{Z^n})] \label{eq:exactresexp}
  \end{equation}
  (where $\bar{P}_{Z^n}\triangleq P_{X^n} \circ W^n$)
  \emph{provided that the limit exists}.
\end{definition}

For the sake of completeness, let us also formally define the error
exponent for an ensemble of random codes.
\begin{definition}\label{def:errorExp}
  Given $\Pi=\{P_{X^n}\in\cP(\cX^n)\}_{n\in\NN}$, a stationary memoryless
  channel $W\colon\cX\to\cY$, and a rate $R$, a number
  $\underline{E}_\rr(\Pi,W,R)$ is called an achievable \emph{error
  exponent} of the ensemble $\Pi$ at rate $R$ on channel $W$, if  
  \begin{equation}
    \liminf_{n\to\infty} -\frac1n \log \E[\Pr\{\hat{s}_{\rm ML}(Y^n) \ne S\}] 
    \ge 
    \underline{E}_\rr(\Pi,W,R)
  \end{equation}
  when $\cC_n$, a random code of size $M=\lceil\exp(nR)\rceil$ is used to
  communicate a uniformly chosen message $S \in\{1,2,\dotsc,M\}$ via $n$
  independent uses of $W$, $y^n$ is the output sequence of $W^n$, and
  $\hat{s}_{\rm ML}(y^n)$ is the ML estimation of $S$ given $y^n$.
\end{definition}
\begin{remark*}
  For the ensembles of interest in this paper, i.e., the ensembles of
  i.i.d.\ and constant-composition random codes the exact error exponents
  are well-known \cite{gallager:it,csiszar:it,gallager:1973}. (The
    exactness of the random exponent of \cite[Theorem~10.2]{csiszar:it}
    follows from exponential tightness of the truncated union bound
  \cite[Appendix~A]{shulman:thesis}.)
\end{remark*}

\begin{definition} \label{def:wiretapCode}
  Given a sequence distributions $\Pi=\{P_{X^n}\in\cP(\cX^n)\}_{n\in\NN}$,
  and a pair of secret message and \emph{random binning} rates $(R_\rs,R)$
  a random \emph{wiretap channel code} is obtained by partitioning a random
  code of size $\lceil\exp[n(R_\rs+R)]\rceil$ in the ensemble of random
  codes defined via $\Pi$ into $M_\rs \doteq \exp(n R_\rs)$ sub-codes (or
  bins) of size $\lfloor\exp(nR)\rfloor$, denoted as
  $\cC_n^s,\,s\in\{1,2,\dotsc,M_\rs\}$, each associated to a message.  To
  communicate the message $s$, the encoder transmits a codeword from the
  sub-code $\cC_n^s$ uniformly at random (thus it requires an entropy rate
  of $R$).  
\end{definition}

\begin{theorem} \label{thm:exactSecExp} 
  Let $W_\rM\colon\cX\to\cY$ and $W_\rE\colon\cX\to\cZ$ be the pair of
  legitimate receiver's and wiretapper's stationary memoryless channels
  respectively (see Figure~\ref{fig:wiretap}).
  Fix a sequence of codeword sampling distributions
  $\Pi=\{P_{X^n}\in\cP(\cX^n)\}_{n\in\NN}$.
  Let $\underline{E}_\rr(\Pi,W_\rM,R)$ be an achievable error exponent
  for the ensemble $\Pi$ over the channel $W_\rM$ at rate $R$ (see
  Definition~\ref{def:errorExp}) and $E_{\rm s}(\Pi,W_\rE,R)$ be the
  \emph{exact} resolvability exponent of the
  ensemble $\Pi$ over the channel $W_\rE$ at rate $R$ (see
  Definition~\ref{def:exactResExp}).  Then
  for any rate pair $(R_\rs,R)$ such that $E_\rs(\Pi,W_\rE,R+R_\rs)
  > E_\rs(\Pi,W_\rE,R)$, using the ensemble of random wiretap channel codes
  constructed as in Definition~\ref{def:wiretapCode}, when the secret
  message $S$ is uniformly distributed,
  \begin{align}
    \liminf_{n\to\infty} -\frac1n \log \E[\Pr\{\hat{s}_{\rm ML}(Y^n)\ne
    S\}] &\ge
    \underline{E}_\rr(\Pi,W_\rM,R+R_\rs) \label{eq:errExp} \\
    \lim_{n\to\infty} -\frac1n \log \E[I(S;Z^n)] 
    &= E_\rs(\Pi,W_\rE,R), \label{eq:secExp}
  \end{align}
  where $\hat{s}_{\rm ML}(y^n)$ is the ML estimation of the sent message given
  $y^n$, the output of legitimate receiver's channel.  In other words,
  $E_\rs$ (evaluated at the random binning rate $R$) is also the
  \emph{exact secrecy exponent} for the ensemble $\Pi$.
\end{theorem}
\begin{IEEEproof}
  That $\underline{E}_\rr(\Pi,W_\rM,R+R_\rs)$ is an achievable error
  exponent for the legitimate receiver is obvious: probability of misdecoding
  the message $S$ is upper-bounded by probability of incorrect decoding of
  the sent codeword. We shall, hence, only prove \eqref{eq:secExp}.

  Since, to communicate a particular message $s \in \cS_n$, the encoder
  transmits a codeword from the code $\cC_n^s$ associated to the message
  $s$, conditioned on $S=s$ the output of $W_\rE^n$ has distribution
  $P_{\cC_n^s}$ and, since $S$ is uniformly distributed, the
  \emph{unconditional} output distribution of $W_\rE^n$ will be $P_{\cC_n}$
  (cf. \eqref{eq:output}).  
  Therefore,  the identity $I(A;B)=D(P_{B|A}\|Q_B|P_A)-D(P_B\|Q_B)$
  (for $(A,B)\sim P_{AB}$ and any arbitrary
  distribution $Q_B$) yields:
  \begin{equation} 
    \E[I(S;Z^n)]  =
    \E[D(P_{\cC_n^S}\|\bar{P}_{Z^n}|P_S)]-\E[D(P_{\cC_n}\|\bar{P}_{Z^n})].  
    \label{eq:secvsres:foo}
  \end{equation}
  Using the linearity of expectation and the fact that the sub-codes
  $\cC_n^s$ are identically distributed we get:
  \begin{align}
    \E[D(P_{\cC_n^S}\|\bar{P}_{Z^n}|P_S)] & = \sum_{s=1}^{M_\rs} P_S(s) 
    \E[D(P_{\cC_n^s}\|\bar{P}_{Z^n})] \nonumber\\
    & = \E[D(P_{\cC_n^1}\|\bar{P}_{Z^n})].
  \end{align}
  Thus, by \eqref{eq:exactresexp}, we have
  \begin{align}
    \lim_{n\to\infty}-\frac1n \log \E[D(P_{\cC_n^s}\|\bar{P}_{Z^n}|P_S)] 
    & = E_\rs(\Pi,W_\rE,R),
    \label{eq:secvsres:bar} \\
    \lim_{n\to\infty}-\frac1n \log\E[D(P_{\cC_n}\|\bar{P}_{Z^n})] & =
    E_\rs(\Pi,W_\rE,R+R_\rs) \nonumber \\
    & > E_\rs(\Pi,W_\rE,R).
    \label{eq:secvsres:baz}
  \end{align}
  where the last inequality follows from the assumption that
  $E_{\rs}(\Pi,W_\rE,R+R_\rs) > E_{\rs}(\Pi,W_\rE,R)$.
  Using \eqref{eq:secvsres:bar} and \eqref{eq:secvsres:baz} in
  \eqref{eq:secvsres:foo} concludes the proof.
\end{IEEEproof}
\begin{remark}   
  That (a lower bound to) the resolvability exponent, lower-bounds the
  secrecy exponent is already used in
  \cite{hayashi:2006,han:2014,bastani:2015}.  Theorem~\ref{thm:exactSecExp}
  complements this result by showing that the exact resolvability exponent
  equals the exact secrecy exponent.
\end{remark}
\begin{remark} 
  To show the achievability of $\underline{E}_\rr$ in the
  proof of Theorem~\ref{thm:exactSecExp}, we used a decoder that estimates
  the sent codeword and then decides to which sub-code it belongs.  In
  \cite{merhav:2014} it has been shown that, when the code sampling
  distribution $P_{X^n}$ depends on $x^n$ only through its type, the error
  exponent of this decoder is the same as that of the \emph{optimal}
  decoder (that computes the likelihood score for each message $s$ by
  summing up the likelihoods of all codewords in $\cC_n^s$ and then decides
  on the most likely message) for an average code in the ensemble.
\end{remark}
\begin{remark}
  Equations~\eqref{eq:errExp} and \eqref{eq:secExp} suggest a trade-off in code
  design in terms of the choice of input distributions,
  $\Pi=\{P_{X^n}\in\cP(\cX^n)\}_{n\in\NN}$.
  The sequence of input distributions $\Pi$ that
  maximizes $E_\rs$ may not coincide with the one that maximizes
  $\underline{E}_\rr$.
\end{remark}

Theorem~\ref{thm:exactSecExp} reduces the problem of computing
the exact secrecy exponent of the ensemble to that of computing the exact
resolvability exponent of the ensemble which is easier as the former
involves the divergence between two random distributions $P_{\cC_n^s}$ and
$P_{\cC_n}$ while the latter depends only on $P_{\cC_n^s}$.  The assumption
on uniform prior of secret messages is crucial to establish such a
result.\footnote{Without such an assumption $I(S;Z^n)=0$,
  namely, the secrecy exponent is infinity if $P_{S}$ is positive only
for a single secret message.}  However, in a practical system, the user
chooses the distribution of the secret messages and it is desirable to have a
worst-case guarantee of performance.  Therefore, before continuing with the
main results of the paper, it is worth mentioning the following result
(which is proved in Appendix~\ref{app:existence}).
\begin{theorem} \label{thm:existence}
  Let $W_\rM\colon\cX\to\cY$ and $W_\rE\colon\cX\to\cZ$ be the pair of
  legitimate receiver's and wiretapper's stationary memoryless channels
  respectively (see Figure~\ref{fig:wiretap}) and 
  $\Pi=\{P_{X^n}\in\cP(\cX^n)\}_{n\in\NN}$ be a sequence of code sampling
  distributions.  If $\underline{E}_\rr(\Pi,W_\rM,R)$ is an achievable
  error exponent for the ensemble $\Pi$ over the channel $W_\rM$ at rate
  $R$ that is continuous in $R$ and $\underline{E}_{\rm s}(\Pi,W_\rE,R)$
  is an achievable resolvability exponent of the ensemble $\Pi$ over the
  channel $W_\rE$, then there exists a sequence of wiretap channel codes of
  secret message $R_\rs$ and random binning rate $R$ in the ensemble
  (indexed by their block-length $n$) using
  which,
  \begin{align}
    \liminf_{n\to\infty} -\frac1n \log \Pr\{\hat{s}_{\rm ML}(Y^n)\ne S\}
    & \ge \underline{E}_\rr(\Pi,W_\rM,R+R_\rs),\\
    \liminf_{n\to\infty} -\frac1n \log I(S;Z^n)
    & \ge \underline{E}_\rs(\Pi,W_\rE,R) 
  \end{align}
  for \emph{any distribution of the secret message $P_S$}.
\end{theorem}
%%%%%%%%%%%%%%%%%%%%%%%%%%%%%%%%%%%%%%%%%%%%%%%%%%%%%%%%%%%%%%%%%%%%%%%%%%%
\section{Exact Resolvability Exponents} \label{sec:exp}
%%%%%%%%%%%%%%%%%%%%%%%%%%%%%%%%%%%%%%%%%%%%%%%%%%%%%%%%%%%%%%%%%%%%%%%%%%%
In light of Theorem~\ref{thm:exactSecExp}, we shall focus on deriving the
exact resolvability exponents for the ensembles of i.i.d.\ and
constant-composition random codes.  Accordingly, $\cC_n$ will denote the
random resolvability code in this section and not the entire wiretap
channel code. 

\subsection{Main Result}
\begin{theorem}\label{thm:finiteLength}
  Let $\cC_n$ be a random code of block-length $n$ and rate $R$ constructed
  by sampling $M=\lfloor\exp(nR)\rfloor$ codewords independently from the
  distribution $P_{X^n}\in\cP(\cX^n)$ (see \eqref{eq:randomcode}). Let
  $W\colon\cX\to\cZ$ be a discrete memoryless channel and $P_{\cC_n}$ be
  the (random) output distribution of $W^n$ when a uniformly chosen
  codeword from $\cC_n$ is transmitted via $n$ independent uses of $W$ (see
  \eqref{eq:output}).  Then,
  \begin{enumerate}[(i)]
    \item
      if $P_{X^n}=P_X^n$ for some $P_X\in\cP(\cX)$, 
      \begin{align} 
	& \E[D(P_{\cC_n}\|\bar{P}_{Z^n})] \nonumber \\
	& \quad \doteq \begin{cases} 
	  \exp\bigl(-n E_{\rs,n}^{\iid}(P_X,W,R)\bigr) & \text{if
	  $I(P_X,W)>0$}, \\
	  0 & \text{if $I(P_X,W)=0$},
	\end{cases}
	\label{eq:iiddoteq}
      \end{align}
      where 
      \begin{subequations} \label{eq:iidexpn}
	\begin{align} 
	  E_{\rs,n}^\iid(P_X,W,R)&=\min_{Q\in\cP_n(\cX\times\cZ)} 
	  \bigl\{D(Q\|P_X\times W) \nonumber\\
	  &\qquad + [R - f(Q\|P_X\times W)]^+\bigr\},
	\end{align}
	with 
	\begin{equation} \label{eq:fdef}
	  f(Q\|Q') \triangleq \sum_{(x,z)\in\cX\times\cZ} Q(x,z) 
	  \log\frac{Q'(x,z)}{Q'_X(x)Q'_Z(z)},
	\end{equation}
	for any two distributions $Q,Q'\in\cP(\cX\times\cZ)$;
      \end{subequations}
    \item
      if $P_{X^n}(x^n) =
      \ind\bigl\{x^n\in\cT_{P_n}^{n}\bigr\}/{\bigabs{\cT_{P_n}^{n}}}$
      for some sequence of $n$-types
      $\{P_n\in\cP_n(\cX)\}_{n\in\NN}$ that
      converge to $P_X\in\cP(\cX)$, i.e.,
      $\lim_{n\to\infty}\abs{P_n-P_X}=0$,
      \begin{align} 
	& \E[D(P_{\cC_n}\|\bar{P}_{Z^n})] \nonumber\\
	& \quad \doteq 
	\begin{cases}
	  \exp \bigl(-n E_{\rs,n}^{\cc}(P_n,W,R)\bigr) & 
	  \text{if $I(P_X,W)>0$},\\
	  0 & \text{if $I(P_X,W)=0$},
	\end{cases}
	\label{eq:ccdoteq}
      \end{align}
      where
      \begin{subequations}  \label{eq:ccexpn}
	\begin{align}
	  E_{\rs,n}^\cc(P_n,W,R) &=
	  \min_{\substack{V\colon\cX\to\cZ:\\P_n \times
	  V\in\cP_n(\cX\times\cZ)}} 
	  \bigl\{D(V\|W|P_n) \nonumber\\
	  & \qquad + [R - g_n(V\|W|P_n)]^+\bigr\},
	\end{align}
	with 
	\begin{align} 
	  g_n(V\|W|P) &\triangleq \omega(V\|W|P) + H(P\circ V)\nonumber\\
	  &\qquad + \min_{\substack{V'\colon\cX\to\cZ: \\P\times
	  V'\in\cP_n(\cX\times\cZ),\\ P\circ V' = P\circ V }}
	  D(V'\|W|P),\label{eq:gndef} 
	\end{align}
	and 
	\begin{equation} \label{eq:omegav}
	  \omega(V\|W|P)\triangleq\sum_{(x,z)\in\cX\times\cZ} P(x)V(z|x)\log
	  W(z|x), 
	\end{equation}
      \end{subequations} 
      for any distribution $P \in \cP(\cX)$ and pair of stochastic matrices
      $V\colon\cX\to\cZ$ and $W\colon\cX\to\cZ$.
  \end{enumerate}
  Recall that in the above $\bar{P}_{Z^n}=P_{X^n}\circ W^n$ (see
  \eqref{eq:ref}).  
\end{theorem}
Theorem~\ref{thm:finiteLength} gives exponentially tight bounds on the
expected divergence between the output distribution of $W^n$, when its
input is a uniformly chosen codeword from a randomly chosen code and the
distribution induced by the code sampling distribution at any finite (but
possibly large) block-length $n$.  As a consequence, the exact exponential
decay rate of the aforementioned divergence, namely the exact resolvability
exponent for the ensembles of interest, is the limit of the exponents of
\eqref{eq:iiddoteq} and \eqref{eq:ccdoteq} as $n$ goes to infinity.  The
exact resolvability exponents have the same forms as \eqref{eq:iidexpn} and
\eqref{eq:ccexpn} except that the search space of the minimizations will
change from the grid of empirical distributions to the set of all
distributions.
\begin{theorem}\label{thm:asymptotic}\
  \begin{enumerate}[(i)]
    \item
      For the sequence of i.i.d.\ random codes of rate $R$, i.e., those
      defined via the sequence of sampling distributions
      $\{P_{X^n}=P_X^n\}_{n\in\NN}$ for some $P_X\in\cP(\cX)$,
      \begin{align}
	& \lim_{n\to\infty} -\frac1n \log(\E[D(P_{\cC_n}\|\bar{P}_{Z^n})])
	\nonumber \\
	& \quad =
	\begin{cases}
	  E_\rs^{\iid}(P_X,W,R) & \text{if $I(P_X,W)>0$},\\
	  +\infty & \text{if $I(P_X,W)=0$},\\
	\end{cases}\label{eq:iidlim}
      \end{align}
      where
      \begin{align}
	E_\rs^\iid(P_X,W,R) &= \min_{Q\in\cP(\cX\times\cZ)}
	\bigl\{D(Q\|P_X\times W) \nonumber \\
	&\qquad + [R - f(Q\|P_X\times W)]^+\bigr\},
        \label{eq:iidexp}
      \end{align}
      and $f$ is defined in \eqref{eq:fdef}.
    \item[(ii)] 
      For the sequence of constant-composition random codes of rate $R$,
      i.e., those defined via the sequence of sampling distributions
      $\bigl\{P_{X^n}=\ind\bigl\{x^n\in\cT_{P_n}^{n}\bigr\}/{\bigabs{\cT_{P_n}^{n}}}\bigr\}_{n\in\NN}$
      for some sequence of $n$-types $\{P_n\in\cP_n(\cX)\}_{n\in\NN}$ that converge to
      $P_X$, namely, $\lim_{n\to\infty} \abs{P_n-P_X} = 0$,
      \begin{align}
	& \lim_{n\to\infty} -\frac1n \log(\E[D(P_{\cC_n}\|\bar{P}_{Z^n})])
	\nonumber \\
	& \quad =
	\begin{cases} 
	  E_\rs^{\cc}(P_X,W,R) & \text{if $I(P_X,W)>0$},\\
	  +\infty & \text{if $I(P_X,W)=0$},
	\end{cases}\label{eq:cclim}
      \end{align}
      where
      \begin{subequations}\label{eq:ccexp}
	\begin{align}
	  E_\rs^\cc(P_X,W,R) 
	  & = \min_{V:\cX\to\cZ}
	  \bigl\{D(V\|W|P_X) \nonumber\\
	  & \qquad + [R - g(V\|W|P_X)]^+\bigr\},
	  \label{eq:ccexpdef}
	\end{align}
	with  
	\begin{align} 
	  g(V\|W|P) & \triangleq \omega(V\|W|P) + H(P\circ V) \nonumber\\
	  & \qquad + \min_{\substack{V':\cX\to\cZ\\ P\circ V' = P\circ V }}
	  D(V\|W|P),
	  \label{eq:gdef}
	\end{align}
      \end{subequations}
      for any distribution $P\in\cP(\cX)$ and pair of stochastic matrices
      $V\colon\cX\to\cZ$ and $W\colon\cX\to\cZ$ (and $\omega$
      defined as in \eqref{eq:omegav}).
  \end{enumerate}
  Both exponents $E_\rs^\iid$ and $E_\rs^\cc$ are positive and strictly
  increasing in $R$ for $R > I(P_X,W)$.  Moreover, the value of
  $E_\rs^\iid$ can be computed through 
  \begin{subequations}\label{eq:iidexpalt}
    \begin{equation} 
      E_\rs^\iid(P_X,W,R) = \max_{0\le\lambda\le1} \{\lambda R -
      F_0(P_X,W,\lambda)\}
    \end{equation}
    with
    \begin{align} 
      &F_0(P_X,W,\lambda)\nonumber\\
      &\quad\triangleq\log\sum_{(x,z)\in\cX\times\cZ} P_X(x) W(z|x)^{1+\lambda}
      (\!P_X\circ W\!)(z)^{-\lambda}.\label{eq:f0} 
    \end{align}
  \end{subequations}
\end{theorem}
Theorem~\ref{thm:asymptotic} is proved in Appendix~\ref{app:lim}.
\begin{corollary} \label{cor:main}
  The exponents $E_\rs^\iid(P_X,W_\rE,R)$ and $E_\rs^\cc(P_X,W_\rE,R)$ of
  \eqref{eq:iidexp} and \eqref{eq:ccexp} are the exact secrecy exponents
  for the ensembles of random wiretap channel codes of rate pair
  $(R,R_\rs)$ constructed from the ensembles of random i.i.d.\ and
  constant-composition codes, respectively, provided that $R_\rs>0$ and
  $R>I(P_X,W_\rE)$.
\end{corollary}
%%%%%%%%%%%%%%%%%%%%%%%%%%%%%%%%%%%%%%%%%%%%%%%%%%%%%%%%%%%%%%%%%%%%%%%%%%%
\subsection{Comparison of Exponents} \label{sec:compare}
Corollary~\ref{cor:main} states that the exponent $E_\rs^\iid$,
which was already derived in \cite{hayashi:2011a,han:2014,bastani:2015} is,
indeed, the exact secrecy exponent for the ensemble of i.i.d.\ random
codes.  (The exponent is expressed in the form of \eqref{eq:iidexpalt} in
\cite{hayashi:2011a,han:2014,bastani:2015}.)  In contrast, it can be shown
that $E_\rs^\cc$, the exact secrecy exponent for the ensemble of
constant-composition random codes, is larger than the previously-derived
lower bound in \cite{hayashi:2011b}:
\begin{subequations} \label{eq:ccweak}
  \begin{equation} \label{eq:ccweakexp}
    \underline{E}_\rs(P_X,W_\rE, R) = \max_{0 \le \lambda \le 1}
    \{\lambda R - E_0(P_X,W_\rE,\lambda)\},
  \end{equation}
  with
  \begin{align} 
    &E_0(P_X,W,\lambda) \nonumber\\
    &\quad\triangleq\log\sum_{z\in\cZ}\Bigl(\sum_{x\in\cX}P_X(x)W(z|x)^{\frac{1}{1-\lambda}}\Bigr)^{1-\lambda}.\label{eq:e0}
  \end{align}
\end{subequations}
(Note that the function $E_0$ in \eqref{eq:e0} is essentially Gallager's $E_0$
\cite{gallager:it} up to a minus sign.)
For every discrete memoryless stationary channel $W\colon\cX\to\cZ$,
\begin{equation} 
  E_\rs^\cc (P_X,W,R) \ge \underline{E}_{\rm
  s}(P_X,W,R).  \label{eq:e0cmp}
\end{equation}
This follows from the fact that $g(V\|W|P)\le I(P,V)$ using
similar steps as in \cite[Problem~10.24]{csiszar:it} to derive
Gallager-style expressions of error exponents (see Appendix~\ref{app:e0cmp}
for a complete proof).

\begin{figure*}[htb]
  \centering
  \subfloat[$P_X(0)=0.3,P_X(1)=0.7$]{\includegraphics{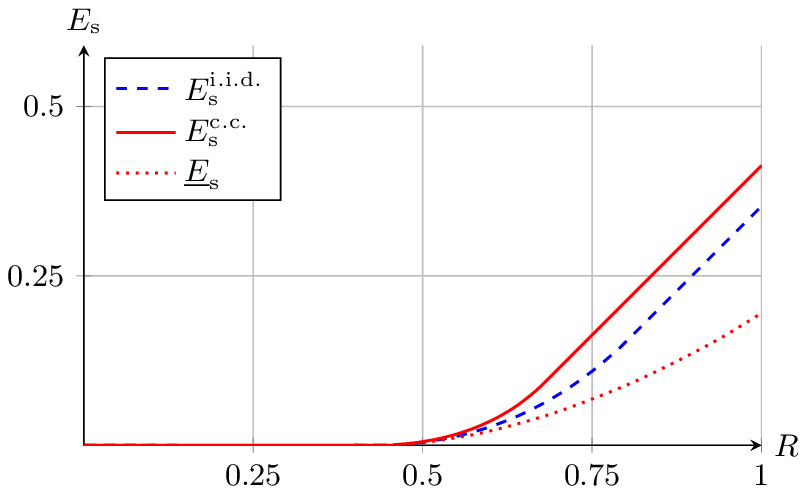}}
  \quad
  \subfloat[$P_X(0)=P_X(1)= 0.5$]{\includegraphics{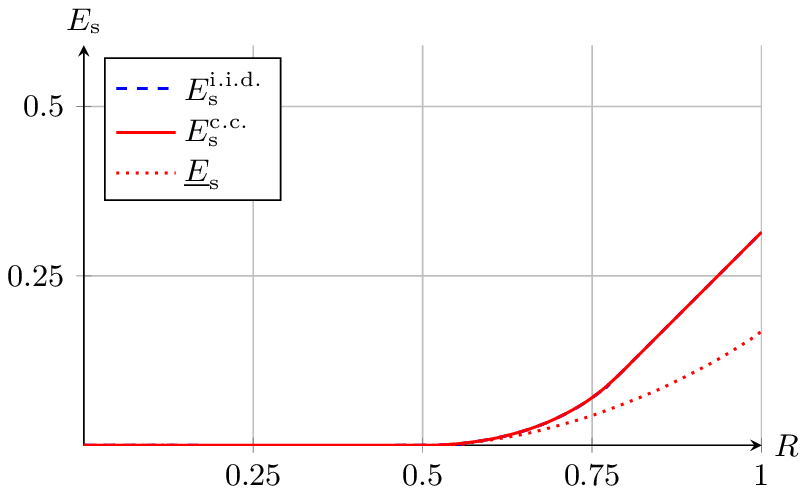}}
  \caption{Comparison of secrecy exponents for Binary Symmetric
  Channel with crossover probability $0.11$}
  \label{fig:bsc}
\end{figure*}

\begin{figure*}[htb]
  \centering
  \subfloat[$P_X(0)=0.28,P_X(1)=0.72$]{\includegraphics{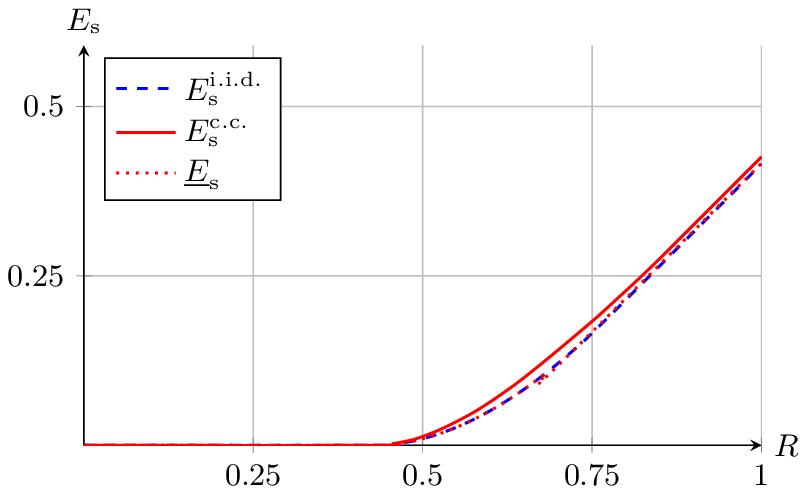}}
  \quad
  \subfloat[$P_X(0)=P_X(1)= 0.5$]{\includegraphics{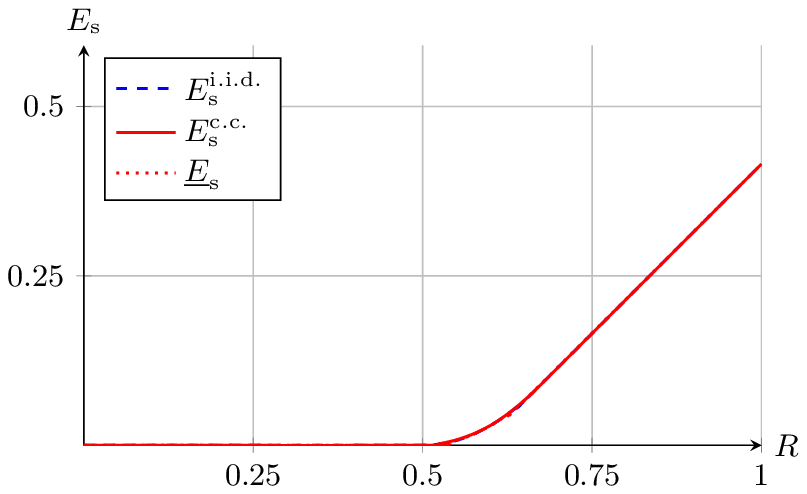}}
  \caption{Comparison of secrecy exponents for Binary Erasure
  Channel with erasure probability $0.5$}
  \label{fig:bec}
\end{figure*}

\begin{figure*}[htb]
  \centering
  \subfloat[$P_X(0)=0.36,P_X(1)=0.64$]{\includegraphics{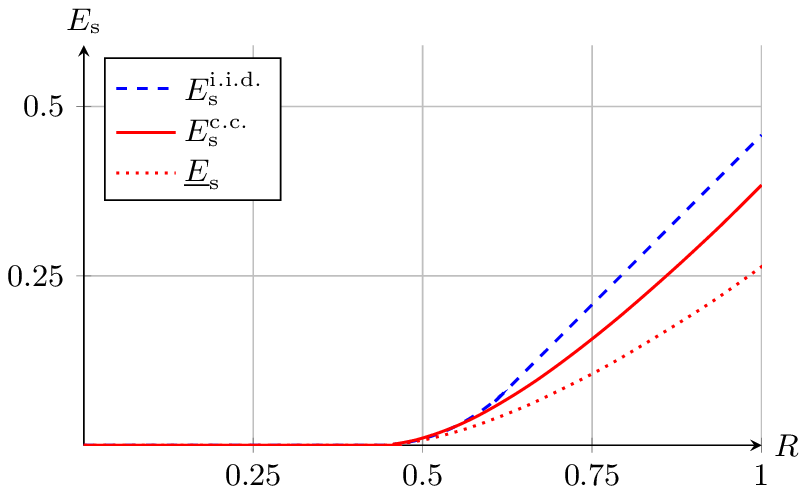}}
  \quad
  \subfloat[$P_X(0)=0.58, P_X(1)=0.42$
  (capacity-achieving)]{\includegraphics{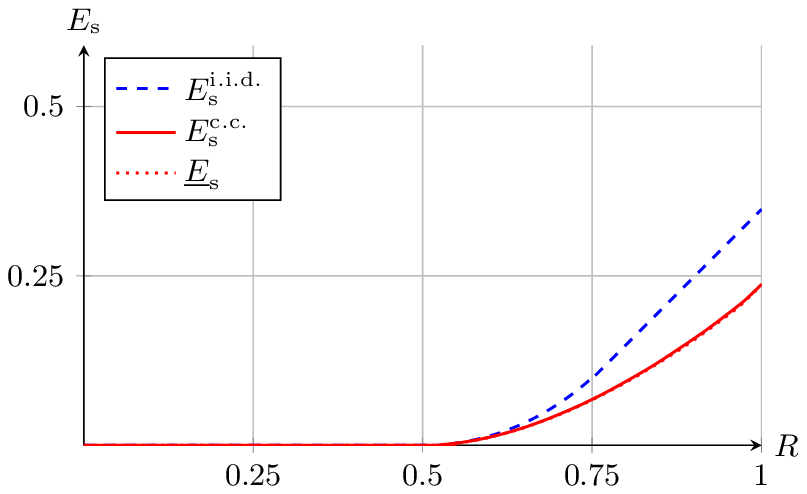}}
  \caption{Comparison of secrecy exponents for Z-channel with
  $W_\rE(0|1) = 0.303$}
  \label{fig:z}
\end{figure*}

\begin{figure*}[htb]
  \centering
  \subfloat[$P_X(0)=0.42,
  P_X(1)=0.58$]{\includegraphics{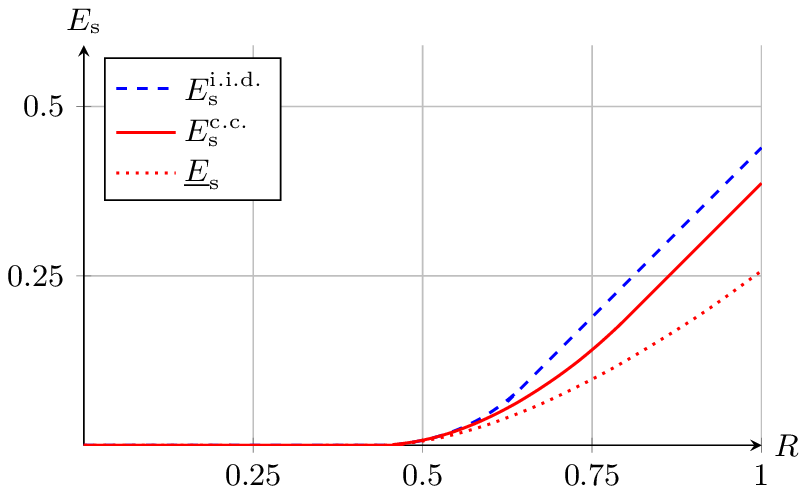}}
  \quad
  \subfloat[$P_X(0)=0.57, P_X(1)=0.43$
  (capacity-achieving)]{\includegraphics{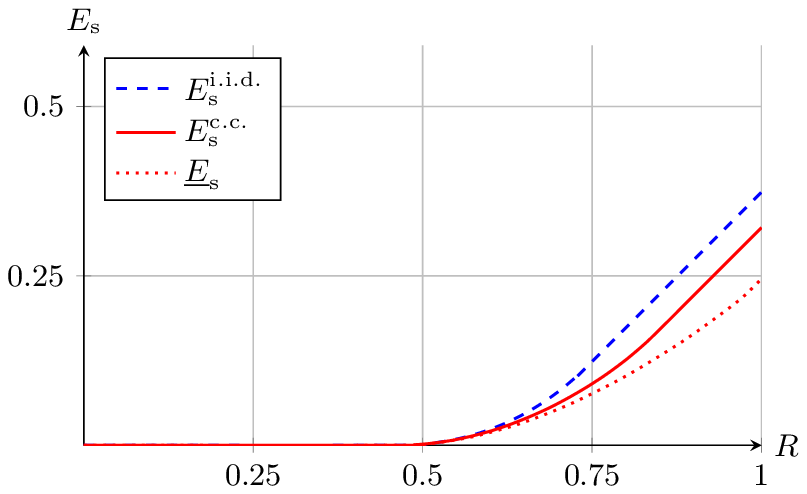}}
  \caption{Comparison of secrecy exponents for binary asymmetric
  channel with $W_\rE(1|0) = 0.01$, $W_\rE(0|1) = 0.303$}
  \label{fig:bac}
\end{figure*}

As for comparing the secrecy exponents $E_\rs^\iid$ and $E_{\rm
s}^\cc$, numerical examples show that in general, there is no
ordering between them.  In particular, as shown in Figures~\ref{fig:bsc}
and \ref{fig:bec}, for the binary symmetric channel and the binary erasure
channel, the ensemble of constant-composition random codes leads to a
larger exponent than the ensemble of i.i.d.\ random codes.  The two
exponents are equal when the input distribution is uniform. On the other
side, in Figures~\ref{fig:z} and \ref{fig:bac}, we see that for asymmetric
channels (the Z-channel and the binary asymmetric channel) the ensemble of
constant-composition random codes results in a smaller secrecy exponent
compared to the ensemble of i.i.d.\ random codes.
The reader may find details on how the exponents are computed in
Appendix~\ref{app:compute}.

%%%%%%%%%%%%%%%%%%%%%%%%%%%%%%%%%%%%%%%%%%%%%%%%%%%%%%%%%%%%%%%%%%%%%%%%%%%
\section{Proof of Theorem~\ref{thm:finiteLength}}\label{sec:proof}
In this section, we fix $P_X$ and set $P_{XZ}(x,z) = P_X(x)
W(z|x)$.  Moreover, we assume, without essential loss of generality, that
(i) $\supp(P_X) = \cX$ (and for the constant-composition codes, $\forall n$,
$\supp(P_n)=\cX$), and (ii) for every $z\in\cZ$, there exists at least one
$x\in\cX$ such that $W(z|x) > 0$.   

Recall that the setting we are considering is as follows:  A random code
$\cC_n = \{X^n_1,\dotsc,X^n_M\}$ of block-length $n$ and size
$M=\lfloor\exp(nR)\rfloor$
is constructed by sampling each codeword independently from 
distribution $P_{X^n}$.  A uniformly chosen codeword from this code is
transmitted through the product channel $W^n$ and the (random)
distribution of its output sequence is as in \eqref{eq:output}. 

%%%%%%%%%%%%%%%%%%%%%%%%%%%%%%%%%%%%%%%%%%%%%%%%%%%%%%%%%%%%
\paragraph*{Trivial Case (zero-capacity channel)}
If $P_X$ is such that $I(X;Z)=0$, then $\forall x\in\cX$ and $\forall
z\in\cZ$, $W(z|x)=P_Z(z)$.    This implies that \emph{for any code
$\cC_n$}, $P_{\cC_n}=P_Z^n$.   Moreover, $\bar{P}_{Z^n} = P_{X^n}\circ W^n
= P_Z^n$ as well, thus, $D(P_{\cC_n}\|\bar{P}_{Z^n}) = 0$ (with probability
$1$ for a random code) which, in turn, implies
$\E[D(P_{\cC_n}\|\bar{P}_{Z^n})]=0$.
% (Note that, in fact, the result in this case holds for any code and any
%   block-length $n$. This is natural---the channel outputs a symbol drawn
%   from the distribution $P_Z$
%   independent of its input, thus as long as the input of $W^n$ is some
%   sequence in $\cX^n$, its output is an i.i.d.\ $P_Z$ sequence of
% length $n$.)
%%%%%%%%%%%%%%%%%%%%%%%%%%%%%%%%%%%%%%%%%%%%%%%%%%%%%%%%%%%%

Now, we begin the non-trivial part of the proof, namely when the channel
output sequence $Z^n$ is correlated with its input.  For any fixed
$z^n\in\cZ^n$, $P_{\cC_n}(z^n)$ is an average of $M$ i.i.d.\
random variables $W^n(z^n|X^n_i)$, $i=1,\dotsc,M$ and, hence, is naturally
expected to concentrate around its mean, which is exactly
$\bar{P}_{Z^n}(z^n)$.  However, since the distribution of each of summands
in \eqref{eq:output} depends on $n$, a plain application of law of large
numbers is not possible in this setting.  
Let
\begin{equation}
  \label{eq:ldef}
  L(z^n) \triangleq \begin{cases}
    \frac{P_{\cC_n}(z^n)}{\bar{P}_{Z^n}(z^n)} & \text{if
    $\bar{P}_{Z^n}(z^n) > 0$},\\
    1 & \text{otherwise},
  \end{cases}
\end{equation}
denote the (random) likelihood ratio of each sequence $z^n \in \cZ^n$. By
construction,
\begin{equation}
  \E[L(z^n)] = 1, \qquad \forall z^n\in\cZ^n.
\end{equation}
Moreover, it follows that $P_{\cC_n}\ll\bar{P}_{Z^n}$ with probability $1$
(see Lemma~\ref{lem:ref}).  Thus, the linearity of expectation yields
\begin{align}
  & \E[D(P_{\cC_n} \| \bar{P}_{Z^n})] 
  = \E\Biggl[ \sum_{z^n \in \cZ^n} P_{\cC_n}(z^n) \log \bigg(
  \frac{P_{\cC_n}(z^n)}{\bar{P}_{Z^n}} \bigg) \Biggr] 
  %\nonumber
  \\
  & \quad = \sum_{z^n \in \cZ^n} \E\biggl[ P_{\cC_n}(z^n) \log \biggl(
  \frac{P_{\cC_n}(z^n)}{\bar{P}_{Z^n}(z^n)} \biggr) \biggr] 
  %\nonumber
  \\
  & \quad = \sum_{z^n \in \cZ^n} \bar{P}_{Z^n}(z^n) \E[L(z^n) \log L(z^n)] 
  \label{eq:divsuml}
  %\nonumber
\end{align}

To prove Theorem~\ref{thm:finiteLength} we derive exponentially tight
bounds on the value of $\E[L(z^n) \log L(z^n)]$ (for each individual
$z^n\in\cZ^n$) and eventually combine those bounds in \eqref{eq:divsuml} to
derive the exponents of Theorem~\ref{thm:finiteLength}.
%%%%%%%%%%%%%%%%%%%%%%%%%%%%%%%%%%%%%%%%%%%%%%%%%%%%%%%%%%%%
\subsection{Preliminaries}
\begin{lemma} \label{lem:ref}
  Let $\bar{P}_{Z^n}$ be as defined in \eqref{eq:ref}.  Then:
  \begin{enumerate}[(i)]
    \item $P_{\cC_n} \ll \bar{P}_{Z^n}$ with probability $1$.
    \item For any codeword sampling distribution $P_{X^n}\in\cP(\cX^n)$
      that depends on $x^n$ only through its type,
      $\bar{P}_{Z^n}(z^n)$ will depend on $z^n$ only through its type.
    \item 
      For both choices of $P_{X^n}$ in Theorem~\ref{thm:finiteLength},
      $\forall z^n\in\supp(\bar{P}_{Z^n})$, $\bar{P}_{Z^n}(z^n) >
      (1/\alpha)^n$ where 
      \begin{equation} 
	\alpha \triangleq \begin{cases}
	  \frac{1}{P_{\min} W_{\min}} & \text{if $P_{X^n}=P_X^n$,}\\
	  \frac{\abs{\cX}}{W_{\min}} & \text{if $P_{X^n} =
	  \frac{\ind\big\{x^n\in\cT_{P_n}^n\bigr\}}{\bigabs{\cT_{P_n}^n}}$},
	\end{cases} 
      \end{equation}
      with 
      $P_{\min} \triangleq \min_{x\in\cX} P_X(x)$ and
      $W_{\min} \triangleq \min_{(x,z)\in\cX\times\cZ\colon W(z|x)>0}W(z|x)$.  
  \end{enumerate}
\end{lemma}
\begin{IEEEproof} 
  See Appendix~\ref{app:ref}.
\end{IEEEproof}
\begin{remark*}
  For the i.i.d.\ random coding ensemble, i.e., when $P_{X^n}=P_X^n$, the
  reference measure $\bar{P}_{Z^n}$ equals the product measure $P_Z^n$ and,
  hence, $\supp(\bar{P}_{Z^n})=\cZ^n$ (since we assumed $\supp(P_X)=\cX$
  and for every $z\in\cZ$ there exists at least one $x\in\cX$ such that
  $W(z|x)>0$).  In contrast, when $P_{X^n}$ is the uniform distribution
  over the type-class $\cT_{P_n}^n$ (i.e., for the constant-composition
  random coding ensemble) the support of $\bar{P}_{Z^n}$ need not
  necessarily be $\cZ^n$. For instance, consider a binary
  erasure channel and $P_n$ being uniform distribution on $\{0,1\}$ (for
  even $n$). Then
  $\bar{P}_{Z^n}$ puts no mass on the all-zero output sequence, and by
  symmetry, neither on the all-one sequence.
\end{remark*}
\begin{lemma} \label{lem:alna}
  Let $A$ be an arbitrary non-negative random variable. Then, for
  any $\theta>0$,
  \begin{equation} \label{eq:alna}
    c(\theta) \Bigl[\frac{\var(A)}{\E[A]} - \tau_\theta(A) \Bigr] 
    \le 
    \E\Bigl[A \ln\Bigl(\frac{A}{\E[A]}\Bigr)\Bigr] 
    \le 
    \frac{\var(A)}{\E[A]}
  \end{equation}
  where 
  \begin{align} \label{eq:taudef}
    \tau_\theta(A) &\triangleq
    \E[A] \Bigl[\theta^2 \Pr\{A>(\theta+1)\E[A]\} \nonumber\\
    &\qquad+
    2\int_{\theta}^{+\infty}v\Pr\{A>(v+1)\E[A]\} \rd v \Bigr],
  \end{align}
  and 
  \begin{equation}  \label{eq:cthetadef}
    c(\theta) \triangleq \frac{(1+\theta)\ln(1+\theta) - \theta}{\theta^2}.
  \end{equation}
\end{lemma}
\begin{IEEEproof} 
  See Appendix~\ref{app:alna}.
\end{IEEEproof}
\begin{remark*}
  It follows from Jensen's inequality that $\E[A\ln(A/\E[A])]) \ge
  0$.  Lemma~\ref{lem:alna} improves this lower bound for random variables
  with sufficiently small tails.
\end{remark*}
Unfortunately, $L(z^n)$ has heavy tails and a direct application of
Lemma~\ref{lem:alna} to $L(z^n)$ will not result in exponentially tight
bounds on $\E\bigl[L(z^n)\log L(z^n)\bigr]$.  However, it turns out that
$L(z^n)$ can be split into light- and heavy-tail components.  As we shall
see shortly, the heavy-tail component contributes to $\E\bigl[L(z^n)\log
L(z^n)\bigr]$ only via its mean and Lemma~\ref{lem:alna} can be applied to
the light-tail component to obtain exponentially tight bounds on
$\E\bigl[L(z^n)\log L(z^n)\bigr]$. 

Since $\bar{P}_{Z^n}(z^n)$ depends on $z^n$ only through its type, we can
use type enumeration method \cite{merhav:2009,merhav:2014} and write
\begin{align}
  L(z^n) & =
  \frac1M \sum_{i=1}^{M} \frac{W^n(z^n|X^n_i)}{\bar{P}_{Z^n}(z^n)} \\
  & = \frac1M \sum_{Q\in\cP_n(\cX\times\cZ)} 
  N_Q(z^n) \ell(Q)
\end{align}
where 
\begin{equation}
  \ell(Q) \triangleq
  \frac{W^n(\tilde{z}^n|\tilde{x}^n)}{\bar{P}_{Z^n}(\tilde{z}^n)} \qquad
  \text{for some $(\tilde{x}^n,\tilde{z}^n)\in\cT_{Q}^n$},
\end{equation}
and
\begin{equation}
  N_Q(z^n) \triangleq \bigabs{\bigl\{x^n \in \cC_n: (x^n,z^n) \in
  \cT^n_Q \bigl\}}
\end{equation}
is the number of codewords in $\cC_n$ that have joint type
$Q$ with $z^n$. Therefore, $\{N_Q(z^n)\colon Q\in\cP_n(\cX\times\cZ)\}$ is a
multinomial collection with cluster size $M$ and success probabilities
\begin{equation} \label{eq:p}
  p_Q(z^n) = \frac{\abs{\cT_Q^n}}{\abs{\cT_{Q_Z}^n}\abs{\cT_{Q_X}^n}}
  P_{X^n}(\cT_{Q_X}^n) \ind\{Q_Z=\type{z^n}\}
\end{equation}
(where $\type{z^n}$ denotes the type of $z^n$) for any code sampling
distribution $P_{X^n}(x^n)$ that depends on $x^n$ through its
type, including our cases of interest. (The above equality is proved in
Appendix~\ref{app:p}.)

Partition $\cP_n(\cX\times\cZ)=\cQ'_n \cup \cQ''_n$ as
\begin{align}
  \cQ_n' & \triangleq \{Q \in \cP_n(\cX\times\cZ): \ell(Q) \le \e^2 M\},  \\
  \cQ_n''& \triangleq \{Q \in \cP_n(\cX\times\cZ): \ell(Q) > \e^2 M\},
\end{align}
and, accordingly, split $L(z^n) = L_1(z^n) + L_2(z^n)$ as 
\begin{align}
  L_1(z^n) &\triangleq \frac{1}{M} \sum_{Q\in\cQ_n'} N_Q(z^n) \ell(Q),\\
  L_2(z^n) &\triangleq \frac{1}{M} \sum_{Q\in\cQ_n''} N_Q(z^n) \ell(Q).
\end{align}
Indeed, $L_1$ turns out to be the light-tail component of $L$ and $L_2$ its
heavy-tail part.  
Let also,
\begin{align}
  \nu(z^n) &\triangleq \var\bigl(L_1(z^n)\bigr) + \frac1M \E[L_1(z^n)]^2,
  \text{ and} \label{eq:nudef}\\
  \mu(z^n) &\triangleq \E[L_2(z^n)].
\end{align}
Using elementary properties of multinomial distribution it can be
verified that 
\begin{subequations}\label{eq:stats}
  \begin{align}
    \nu(z^n)&=\frac1M \sum_{Q \in \cQ'_n}
    \ell(Q)^2 p_Q(z^n) \label{eq:nuval}\\
    \mu(z^n) &= \sum_{Q \in \cQ''_n} \ell(Q)
    p_Q(z^n) \label{eq:muval}
  \end{align}
\end{subequations}
(A proof of the above is given in Appendix~\ref{app:sumstat} for
completeness.)
In the following two subsections we prove that $\forall
z^n\in\supp(\bar{P}_{Z^n})$, 
\begin{equation} \label{eq:typedot}
  \E\bigl[L(z^n) \ln L(z^n) \bigr] + \frac{1}{M} \doteq 
  \nu(z^n) + \mu(z^n). 
\end{equation}
Since $z^n$ is fixed in both sides of \eqref{eq:typedot} we drop it
in subsections~\ref{sec:proof:direct} and \ref{sec:proof:converse}
to avoid cumbersome notation.
%%%%%%%%%%%%%%%%%%%%%%%%%%%%%%%%%%%%%%%%%%%%%%%%%%%%%%%%%%%%%%%%%%%%%%%%%%%
\subsection{Achievability}\label{sec:proof:direct}
For non-negative $l_1$ and $l_2$, and $l =  l_1+l_2$, 
\begin{align} 
  l \ln (l)  & = l_1 \ln (l) + l_2 \ln (l)  \\
  & = l_1 \ln (l_1) + l_1 \ln (1 + l_2/l_1) + l_2 \ln (l) \\
  & \le l_1 \ln (l_1) + l_2 (1 + \ln(l))
\end{align}
(since $\ln(1+l_2/l_1) \le l_2/l_1$), thus,
\begin{align}
  \E[L\ln L] &\le \E[L_1 \ln L_1] + \E[L_2(1+\ln L)]
  \\
  & \stackrel{(\ast)}{\le} \E[L_1\ln L_1] +(1+n
  \ln\alpha) \E[L_2]
  \label{eq:upper:expsplit}
\end{align}
where $(\ast)$ follows from (iii) in Lemma~\ref{lem:ref}
(as $L=L(z^n)\le1/\bar{P}_{Z^n}(z^n)$).
The upper bound of \eqref{eq:alna} implies
\begin{equation} 
  \E[L_1\ln L_1]\le\E[L_1]\ln\bigl(\E[L_1]\bigr)+\frac{\var(L_1)}{\E[L_1]}
  \stackrel{(\ast)}{\le}\frac{\var(L_1)}{\E[L_1]}
  \label{eq:upper:slogs1}
\end{equation}
where $(\ast)$ follows since $\E[L_1]\le\E[L]=1$.  Moreover, using
\eqref{eq:nudef} and the fact that $\E[L_1]+\E[L_2]=1$ we have
\begin{align}
  \frac{\var(L_1)}{\E[L_1]} &= \frac{\nu}{\E[L_1]}-\frac{\E[L_1]}{M}\\
  &= \nu\Bigl(1+\frac{\E[L_2]}{\E[L_1]}\Bigr) - \frac{1-\E[L_2]}{M}\\
  &= \nu + \E[L_2]\Bigl(\frac{\nu}{\E[L_1]}+\frac1M\Bigr) - \frac1M.
  \label{eq:upper:vars}
\end{align}
Since $\ell(Q)\le M\e^2$ for $Q\in\cQ'_n$, using \eqref{eq:nuval} we have
\begin{equation}\label{eq:nubound}
  \nu \le \frac{1}{M} \sum_{Q\in\cQ'_n} \e^2 M \cdot \ell(Q) p_Q  = 
  \e^2 \E[L_1].
\end{equation}
Using the above in \eqref{eq:upper:vars} and replacing $\E[L_2]=\mu$, we
get
\begin{equation}
  \frac{\var(L_1)}{\E[L_1]} + \frac{1}{M} \le \nu +
  \E[L_2]\Bigl(\e^2+\frac1M\Bigr) \le \nu + (1+\e^2) \mu,
  \label{eq:upper:vars2}
\end{equation}
(since $M \ge 1$).
Finally, using \eqref{eq:upper:vars2} in \eqref{eq:upper:slogs1} yields,
\begin{equation}
  \EE[L_1 \ln L_1] + \frac1M \dotle \nu + \mu.
  \label{eq:upper:slogs2}
\end{equation} 
Using \eqref{eq:upper:slogs2} in \eqref{eq:upper:expsplit} (and noting that
$\alpha \ge 1$ only depends on $\abs{\cX}$,  $P_X$, and  $W$) we conclude that
\begin{equation} \label{eq:upper:sum}
  \E[L \ln L] +\frac{1}{M} \dotle
  \nu + \mu.
\end{equation}
%%%%%%%%%%%%%%%%%%%%%%%%%%%%%%%%%%%%%%%%%%%%%%%%%%%%%%%%%%%%
\subsection{Ensemble Converse}\label{sec:proof:converse}
The choice of $\cQ''_n$ implies
\begin{equation} \label{eq:tbound} 
  \Pr\bigl\{L_2 \in (0,\e^2)\bigr\} = 0.
\end{equation}
This holds since either $\forall Q\in\cQ''_n\colon N_Q=0$ which implies
$L_2=0$ or $\exists Q_0\in\cQ''_n$ such that $N_{Q_0}\ge1$, in which case,
\begin{equation}
  L_2 \ge \frac1M \ell(Q_0) N_{Q_0}  
  \ge \frac1M{\ell(Q_0)} \ge \e^2, \end{equation}
(because $\forall Q\in\cQ''_n$, $\ell(Q) > \e^2 M$).
Consequently,
\begin{align}
  \E[L_2 \ln L_2] & = \sum_{l \ge \e^2} l \, \ln(l) \Pr\{L_2=l\} \\
  & \ge \ln(\e^2) \sum_{l \ge \e^2} l \Pr\{L_2=l\} = 2\E[L_2].
  \label{eq:tlnt}
\end{align}

For positive $l_1$ and $l_2$, and $l=l_1+l_2\ge\max\{l_1,l_2\}$, 
\begin{align}
  l \ln(l) & = l_1 \ln(l) + l_2 \ln(l) \\
  & \ge l_1 \ln (l_1) + l_2 \ln(l_2). \label{eq:lower:ulnu}
\end{align}
Therefore,
\begin{equation}
  \E[L \ln L] \ge \E[L_1 \ln L_1] + \E[L_2 \ln L_2].
\end{equation}
Using the lower bound of \eqref{eq:alna} (with $\tau_\theta(L_1)$ and
$c(\theta)$ defined as in \eqref{eq:taudef} and \eqref{eq:cthetadef}
respectively), $\forall \theta>0\colon$
\begin{align}
  & \E[L_1 \ln L_1] \ge \E[L_1] \ln(\E[L_1]) + c(\theta)
  \Bigl[\frac{\var(L_1)}{\E[L_1]} - \tau_{\theta}(L_1) \Bigr]\\
  & \quad \aeq (1-\E[L_2]) \ln(1-\E[L_2]) + c(\theta)
  \Bigl[\frac{\var(L_1)}{\E[L_1]} - \tau_{\theta}(L_1) \Bigr]\\
  & \quad \bge -\E[L_2] + c(\theta)
  \Bigl[\frac{\var(L_1)}{\E[L_1]} - \tau_{\theta}(L_1)\Bigr]. 
  \label{eq:lower:slns}
\end{align}
In the above (a) follows since $\E[L_1]=1-\E[L_2]$
and (b) since $(1-\varepsilon) \ln(1-\varepsilon) \ge -\varepsilon$.
Using \eqref{eq:tlnt}
and \eqref{eq:lower:slns} in \eqref{eq:lower:ulnu} shows that 
$\forall\theta>0\colon$
\begin{equation}
  \E[L \ln L] \ge
  c(\theta) \Bigl[\frac{\var(L_1)}{\E[L_1]} - \tau_\theta(L_1) \Bigr]
  + \E[L_2].
  \label{eq:lower:split}
\end{equation}

Now we shall upper-bound $\tau_{\theta}(L_1)$.  Starting by bounding the
tail of $L_1$ we have
\begin{align}
  & \Pr\{L_1 \ge (v+1) \E[L_1]\} \nonumber \\
  & \quad = \Pr\left\{
  \sum_{Q\in\cQ'_n} \ell(Q) (N_Q - M p_Q)
  \ge M v \E[L_1]   \right\} \\
  & \quad \le \Pr\left\{ \bigcup_{Q\in\cQ'_n} \left\{%
      \ell(Q)(N_Q-Mp_Q) \ge% 
      \frac{M v \E[L_1]}{\abs{\cQ'_n}}%
    \right\}%
  \right\} \\
  & \quad \ale
  \sum_{Q\in\cQ'_n} 
  \Pr\left\{%
    \ell(Q)(N_Q-Mp_Q)%
    \ge \frac{M v \E[L_1]}{\abs{\cQ'_n}}%
  \right\} \\
  & \quad \ble 
  \sum_{Q\in\cQ'_n}
  \frac{\E[\ell(Q)^4(N_Q-M p_Q)^4]}
  {(M v \E[L_1]/\abs{\cQ'_n})^4} \\
  & \quad =
  \frac{\abs{\cQ'_n}^4}{v^4(\E[L_1])^4} \frac{1}{M^4} \sum_{Q\in\cQ'_n}
  \ell(Q)^4 \E[(N_Q - M p_Q)^4], \label{eq:stail:sum}
\end{align}
where (a) is the union bound and (b) follows by Markov inequality.  
For $N \sim \mathrm{Binomial}(M,p)$, 
\begin{align}
  \E[(N-M p)^4] &= Mp(1-p)[1 + 3(M-2)p(1-p)] \\
  & \le\var(N)+3\var(N)^2.
\end{align} 
Continuing \eqref{eq:stail:sum} we have
\begin{align}
  & \frac{1}{M^4} \sum_{Q\in\cQ'_n} \ell(Q)^4 \E[(N_Q - M p_Q)^4]
  \nonumber\\
  & \quad \le 
  \frac{1}{M^4} \sum_{Q\in\cQ'_n} \ell(Q)^4 
  \bigl(\var(N_Q)+3\var(N_Q)^2\bigr) \\
  & \quad \adotle
  \frac{1}{M^2} \sum_{Q\in\cQ'_n} \ell(Q)^2
  \var(N_Q) + 3 \frac{1}{M^4} \sum_{Q\in\cQ'_n} 
  \ell(Q)^4 \var(N_Q)^2 \\
  & \quad \ble
  \frac{1}{M^2} \sum_{Q\in\cQ'_n} \ell(Q)^2
  \var(N_Q) \nonumber \\
  & \quad \qquad + 3 \Bigl[\frac{1}{M^2} \sum_{Q\in\cQ'_n} \ell(Q)^2 \var(N_Q)
  \Bigr]^2 \\
  & \quad \cle 
  \nu + 3 \nu^2 \ddoteq \nu, \label{eq:stail:nu}
\end{align} 
where (a) follows since $\ell(Q) \le \e^2 M \doteq M$ for
$Q\in\cQ'_n$, (b) since for positive summands, the sum of the squares is
less than the square of the sums, (c) since $\var(N_Q) \le M p_Q$, and (d)
since $\nu\le\e^2\E[L_1]\le\e^2$ (see \eqref{eq:nubound}).  Plugging
\eqref{eq:stail:nu} into \eqref{eq:stail:sum} we get
\begin{equation}
  \Pr\{L_1\ge(v+1)\EE[L_1]\} \dotle \frac{\abs{\cQ'_n}^4\nu}{(\EE[L_1])^4}
  \cdot
  \frac{1}{v^4}.
\end{equation}
Using the above in \eqref{eq:taudef} we get
\begin{align}
  \tau_{\theta}(L_1) & = \E[L_1] \Bigl[\theta^2
    \Pr\{L_1>(\theta+1)\E[L_1]\} \nonumber \\
    & \quad + 
    2 \int_{\theta}^{+\infty} v \Pr\{L_1> (v+1)\E[L_1]\} \rd v
  \Bigr]
  \\
  & \dotle \E[L_1] \Bigl[
    \frac{\theta^2}{\theta^4}
    +
    2 \int_{\theta}^{+\infty} \frac{v}{v^4} \rd v
  \Bigr]
  \frac{\abs{\cQ'_n}^4}{\E[L_1]^4} \nu    \\
  & \doteq \frac{\nu}{\E[L_1]^3} \cdot \frac{\abs{\cQ'_n}^4}{\theta^2}.
  \label{eq:tausbound}
\end{align}
Since \eqref{eq:tausbound} implies $\tau_{\theta}(L_1) \le d(n)
\abs{\cQ'_n}^4 \nu/\bigl(\theta^2 \E[L_1]^3\bigr)$ for some sub-exponentially
increasing sequence $d(n)$ (which only depends on $\abs{\cX}$ and
$\abs{\cZ}$),
taking
\begin{equation} \label{eq:theta}
  \theta_n \triangleq 2 \sqrt{d(n)} \frac{\abs{\cQ'_n}^2}{\E[L_1]},
\end{equation} 
we will have 
\begin{equation}
  \tau_{\theta_n}(L_1) \le \frac14 \cdot \frac{\nu}{\E[L_1]}.
  \label{eq:taus}
\end{equation}
Using \eqref{eq:nudef} and \eqref{eq:taus} in \eqref{eq:lower:split}
we have 
\begin{align}
  & \E[L(z^n)\ln L(z^n)] \ge
  c(\theta_n)\Bigl[\frac{\var(L_1)}{\E[L_1]}-\tau_{\theta_n}(L_1)\Bigr] + \E[L_2] \\
  & \quad \ge c(\theta_n) \Bigl[\frac{\nu}{\E[L_1]}-\frac{1}{M}\E[L_1]-\frac14
  \cdot
  \frac{\nu}{\E[L_1]}\Bigr] + \E[L_2] \\
  & \quad \stackrel{(\ast)}{\ge} 
  c(\theta_n) \Bigl[\frac34\cdot\frac{\nu}{\E[L_1]} -  \frac{1}{M}\Bigr] +
  \E[L_2]
  \label{eq:lower:split2}
\end{align}
(where $(\ast)$ follows because $\E[L_1] \le 1$).  
Since for $\theta>0$, $c(\theta)\le c(0) = \frac12 < 1$, we can further
lower-bound \eqref{eq:lower:split2} as 
\begin{equation}
  \E[L \ln L] \ge \frac34 c(\theta_n)\frac{\nu}{\E[L_1]} + \E[L_2]
  - \frac{1}{M} \label{eq:lower:split3}
\end{equation}
Moreover,
\begin{align}
  c(\theta_n) & = \frac1{\theta_n} \cdot
  \frac{(1+\theta_n)\ln(1+\theta_n)-\theta_n}{\theta_n} \\
  & \age \frac{1}{\theta_n} \cdot 
  \frac{(1+\E[L_1]\theta_n)\ln(1+\E[L_1]\theta_n)-\E[L_1]\theta_n}{\E[L_1]\theta_n} \\
  & = \E[L_1] 
  \frac{(1+\E[L_1]\theta_n)\ln(1+\E[L_1]\theta_n)-\E[L_1]\theta_n}{(\E[L_1]\theta_n)^2}
  \\
  & \bdotge \E[L_1], \label{eq:lower:c}
\end{align}
where (a) follows since $\frac{(1+\theta)\ln(1+\theta)-\theta}{\theta}$
is increasing in $\theta$ and $\E[L_1]\le1$, and (b) since
$\frac{(1+\theta)\ln(1+\theta)-\theta}{\theta^2}$ is decreasing in
$\theta$ (see Lemma~\ref{lem:incdec} in Appendix~\ref{app:alna}) and
$\E[L_1]\theta_n=2\sqrt{d(n)}\abs{\cQ'_n}^2\le2\sqrt{d(n)}(n+1)^{2\abs{\cX}\abs{\cZ}}$.  
Using this lower bound in \eqref{eq:lower:split3} we get
\begin{equation} \label{eq:lower:sum}
  \E[L \ln L] + \frac{1}{M} \dotge \nu + \mu
\end{equation}
%%%%%%%%%%%%%%%%%%%%%%%%%%%%%%%%%%%%%%%%%%%%%%%%%%%%%%%%%%%%
\subsection{Derivation of Exponents for Each Ensemble}
Equations~\eqref{eq:upper:sum} and \eqref{eq:lower:sum} prove
\eqref{eq:typedot}.  Plugging in the values of $\nu(z^n)$ and $\mu(z^n)$
from \eqref{eq:nuval} and \eqref{eq:muval} and continuing
\eqref{eq:typedot}, we get
\begin{align}
  & \E\bigl[L(z^n)\ln L(z^n) \bigr] + \frac1M \doteq 
  \nu(z^n) + \mu(z^n) \\
  & \quad = \sum_{Q\in\cP_n(\cX\times\cZ)} \ell(Q) p_Q(z^n)
  \kappa\bigl(\ell(Q)/M\bigr)
  \label{eq:typedot2}
\end{align}
where 
\begin{equation}
  \kappa(\lambda) = \begin{cases}
    1 & \lambda > \e^2, \\
    \lambda & \lambda \le \e^2. 
  \end{cases}
\end{equation}
It is easy to check that
\begin{equation}
  \min\{1,\lambda\} \le \kappa(\lambda) \le \e^2 \min\{1,\lambda\}
\end{equation}
Therefore, \eqref{eq:typedot2} can be simplified as
\begin{align}
  &\E\bigl[L(z^n)\ln L(z^n) \bigr] + \frac1M \nonumber\\
  & \quad \doteq 
  \sum_{Q\in\cP_n(\cX\times\cZ)} \ell(Q) p_{Q}(z^n)
  \min\Bigl\{1,\frac{\ell(Q)}{M}\Bigr\}.
\end{align}
Using the above in \eqref{eq:divsuml} we get
\begin{align}
  & \E[D(P_{\cC_n}\|\bar{P}_{Z^n})] + \frac{\log(\e)}{M}\nonumber\\
  & \quad \doteq 
  \sum_{z^n \in \cZ^n} \bar{P}_{Z^n}(z^n)
  \sum_{Q\in\cP_n(\cX\times\cZ)} \ell(Q) p_{Q}(z^n)
  \min\Bigl\{1,\frac{\ell(Q)}{M}\Bigr\} \\
  & \quad = 
  \sum_{Q\in\cP_n(\cX\times\cZ)} \ell(Q) 
  \min\Bigl\{1,\frac{\ell(Q)}{M}\Bigr\} 
  \sum_{z^n\in\cZ^n} p_{Q}(z^n) \bar{P}_{Z^n}(z^n).
\end{align}
Plugging in the value of $p_{Q}(z^n)$ from \eqref{eq:p} we get
\begin{equation}\label{eq:exp:sumz}
  \sum_{z^n\in\cZ^n} p_{Q}(z^n) \bar{P}_{Z^n} (z^n) = 
  \frac{\abs{\cT_Q^n}}{\abs{\cT_{Q_X}^n}
  \abs{\cT_{Q_Z}^n}} P_{X^n}\bigl(\cT_{Q_X}^n\bigr)
  \bar{P}_{Z^n}\bigl(\cT_{Q_Z}^n\bigr).
\end{equation}
Moreover, defining
\begin{equation}
  \omega(Q) = \sum_{x,z} Q(x,z) \log W(z|x),
\end{equation}
and recalling that $\bar{P}_{Z^n}$ depends on $z^n$ only through its type,
we deduce that
\begin{equation} \label{eq:exp:lq}
  \ell(Q) =
  \frac{\exp\bigl(n\omega(Q)\bigr)}{\bar{P}_{Z^n}\bigl(\cT_{Q_Z}^n\bigr)/
  \abs{\cT_{Q_Z}^n}}
\end{equation}
Combining \eqref{eq:exp:sumz} and \eqref{eq:exp:lq} yields
\begin{align}
  & \ell(Q) \sum_{z^n} p_{Q}(z^n) \bar{P}_{Z^n}(z^n) =
  \exp\bigl\{n \omega(Q)\bigr\} \abs{\cT_{Q}^n}
  \frac{P_{X^n}\bigl(\cT^n_{Q_X}\bigr)}{\bigabs{\cT_{Q_X}^n}}
  \\
  & \quad \doteq \exp\bigl\{ -n D(Q\|Q_X\times W)\bigr\}
  P_{X^n}\bigl(\cT_{Q_X}^n\bigr),
\end{align}
where the last equality follows since $|\cT_Q^n| \doteq \exp\{n H(Q)\}$
(respectively, $|\cT_{Q_X}^n|\doteq\exp\{nH(Q_X)\}$).
Thus, we have
\begin{align}
  &\E[D(P_{\cC_n}\|\bar{P}_{Z^n})] + \frac{\log(\e)}{M} \nonumber\\
  & \quad \doteq 
  \sum_{Q\in\cP_n(\cX\times\cZ)} \exp\bigl\{-nD(Q\|Q_X\times W)\bigr\}
  \nonumber \\
  & \quad \phantom{\doteq\sum_{Q\in\cP_n(\cX\times\cZ)}}
  \times P_{X^n}\bigl(\cT_{Q_X}^n\bigr) 
  \min\Bigl\{1,\frac{\ell(Q)}{M}\Bigr\}.
  \label{eq:divdot1}
\end{align}
Observe that since 
\begin{equation}
  \ell(P_{XZ}) \ge \exp\{n\omega(P_{XZ})\}\bigabs{\cT_{P_Z}^n} \dotge
  \exp\{n I(X;Z)\},
\end{equation}
taking $Q=P_{XZ}$ shows that the right-hand-side of \eqref{eq:divdot1}
decays at most as fast as $\exp\{-n [R-I(X;Z)]^+\}$ which is
strictly slower than $\frac1M=\exp(-n R)$ since $I(X;Z) > 0$.
Consequently we can ignore the term $\frac{\log(\e)}{M}$ on the
left-hand-side of \eqref{eq:divdot1} and conclude that
\begin{align}
  \E[D(P_{\cC_n}\|\bar{P}_{Z^n})]
  & \doteq 
  \sum_{Q\in\cP_n(\cX\times\cZ)}  \exp\bigl\{-nD(Q\|Q_X\times W)\bigr\}
  \nonumber \\
  & \qquad
  \times P_{X^n}\bigl(\cT_{Q_X}^n\bigr) 
  \min\Bigl\{1,\frac{\ell(Q)}{M}\Bigr\}.
  \label{eq:divdot}
\end{align}
(The careful reader may argue that $P_{XZ}$ may not be an
$n$-type for all $n$ and, hence, find our reasoning for the passage from
\eqref{eq:divdot1} to \eqref{eq:divdot} inaccurate.  While this concern
is valid, the claim is true regardless as we can always find a sequence of
$n$-types that converge to $P_{XZ}$.  We give a rigorous and more detailed
proof of \eqref{eq:divdot} in Appendix~\ref{app:divdot}.)
\subsubsection{Ensemble of i.i.d.\ random codes}
When $P_{X^n}=P_X^n$, 
\begin{equation} \label{eq:iidpx}
  P_{X^n}(\cT_{Q_X}^n)\doteq\exp\{-nD(Q_X\|P_X)\}
\end{equation}
Moreover, $\bar{P}_{Z^n}(z^n)=P_Z^n(z^n)$ (where $P_Z=P_X\circ W$).
Therefore, $\bar{P}_{Z^n}(z^n) = \exp\{n \sum_{z} Q_Z(z) \log P_Z(z)\}$ if
$z^n\in\cT_{Q_Z}^n$.  Therefore,
\begin{align}
  \ell(Q) = \frac{\exp\{n \omega(Q)\}}{P_Z^n(z^n)} & = 
  \exp\Bigl\{n \sum_{x,z} Q(x,z) \log\frac{W(z|x)}{P_Z(z)}\Bigr\}
  \nonumber\\
  & = \exp\bigl\{n f(Q\|P_{XZ})\bigr\}.
\end{align}
where $f$ is defined in \eqref{eq:fdef}.  As a consequence,
\begin{equation} \label{eq:iidmin}
  \min\{1,\ell(Q)/M\} \doteq \exp\bigl\{-n [R - f(Q\|P_{XZ})]^+\bigr\}.
\end{equation}
Using \eqref{eq:iidpx} and \eqref{eq:iidmin} in \eqref{eq:divdot} (together
with the fact that $|\cP_n(\cX\times\cZ)|\le(n+1)^{|\cX||\cZ|}$)
conclude that 
\begin{align}
  & \E[D(P_{\cC_n}\|\bar{P}_{Z^n})] \doteq  \exp\Bigl\{- n \min_{Q\in\cP_n(\cX\times\cZ)} \bigl\{
  D(Q\|Q_X\times W)\nonumber \\
  & \quad\qquad +D(Q_X\|P_X)+[R-f(Q\|P_{XZ})]^+\bigr\} \Bigr\}.
\end{align}
Simplifying the above exponent yields \eqref{eq:iidexpn}.
%%%%%%%%%%%%%%%%%%%%%%%%%%%%%%%%%%%%%%%%%%%%%
\subsubsection{Ensemble of constant-composition random codes}
When the code sampling distribution, $P_{X^n}$, is the uniform distribution
over the type-class $\cT_{P_n}^n$, $P_{X^n}\bigl(\cT_{Q_X}^n\bigr)=0$ unless
$Q_X=P_n$, i.e., $Q = P_n \times V$ for some $V\colon\cX\to\cZ$ such that
$P_n \times V\in\cP_n(\cX\times\cZ)$. (To keep the notation simple, we omit
this last condition from the following equations.)  Therefore
\eqref{eq:divdot} reduces to 
\begin{align} 
  \E[D(P_{\cC_n}\|\bar{P}_{Z^n})]
  &\doteq
  \sum_{V\colon\cX\to\cZ} \exp\bigl\{-nD(V\|W|P_n)\bigr\}\nonumber\\ 
  &\qquad\times\min\{1,\ell(P_n\times V)/M\}.
  \label{eq:ccdivdot}
\end{align}
It remains to evaluate 
\begin{equation}
  \ell(P_n\times V) =
  \frac{W^n(z^n|x^n)}{\bar{P}_{Z^n}(z^n)},
\end{equation}
for some $x^n\in\cT_{P_n}^n$ and $z^n\in\cT^n_V(x^n)$,
where $\cT^n_V(x^n)$ is the $V$-shell of $x^n$.  To this end, we note that
\begin{align}
  &\bar{P}_{Z^n}(z^n) = \frac{1}{|\cT_{P_n}^n|} \sum_{x^n\in\cT_{P_n}^n}
  W(z^n|x^n) \\
  & \quad = \frac{1}{|\cT_{P_n}^n|} \sum_{x^n\in\cT_{P_n}^n} W(z^n|x^n)
  \sum_{V'\colon\cX\to\cZ} \ind\{z^n\in\cT_{V'}^n(x^n)\} \\
  &\quad= \frac{1}{|\cT_{P_n}^n|} \sum_{x^n\in\cT_{P_n}^n}
  \sum_{V'\colon\cX\to\cZ} \ind\{z^n\in\cT_{V'}^n(x^n)\}
  W(z^n|x^n)  \\ 
  & \quad = \frac{1}{|\cT_{P_n}^n|}
  \sum_{x^n\in\cT_{P_n}^n}
  \sum_{V'\colon\cX\to\cZ} \ind\{z^n\in\cT_{V'}^n(x^n)\} \nonumber\\
  & \quad\qquad \times \exp\bigl[-n(D(V'\|W|P_n)+H(V'|P_n))\bigr]  \\
  & \quad = 
  \sum_{V'\colon\cX\to\cZ} 
  \frac{1}{|\cT_{P_n}^n|} \sum_{x^n\in\cT_{P_n}^n}
  \ind\{z^n\in\cT_{V'}^n(x^n)\} \nonumber \\
  & \quad\qquad \times 
  \exp\bigl[-n(D(V'\|W|P_n)+H(V'|P_n))\bigr].
  \label{eq:refcc:foo}
\end{align}
(Recall again that $V'$ must also be such that $P_n\times V'$
  is an $n$-type but we omit this condition from the equations for the sake
of brevity.)
As we have already
shown in the proof of \eqref{eq:p} (cf.\ Appendix~\ref{app:p}),
\begin{align}
  & \frac{1}{\abs{\cT_{P_n}^n}} 
  \sum_{x^n\in\cT_{P_n}^n} \ind\{z^n\in\cT_{V'}^n(x^n)\} \nonumber\\
  & \quad =
  \frac{\abs{\cT_{P_n\times V'}^n}}{\abs{\cT^n_{P_n}}\abs{\cT^n_{P_n\circ
  V'}}}
  \ind\{P_n\circ V' = \type{z^n}\}\\ 
  & \quad \doteq 
  \exp\bigl[n[H(V'|P_n)-H(P_n\circ V')]\bigr] \ind\{P_n\circ
  V'=\type{z^n}\}
  \label{eq:refcc:bar}
\end{align}
(where $\type{z^n}$ is the type of $z^n$).  Using
\eqref{eq:refcc:bar} in \eqref{eq:refcc:foo} and recalling that
$z^n$ has type $P_n\circ V$ we get
\begin{align}
  \bar{P}_{Z^n}(z^n) & \doteq \exp\biggl[-n [H(P_n\circ V) \nonumber \\
    & \qquad +
      \min_{\substack{V'\colon\cX\to\cZ\\P_n\circ V=P_n\circ V}}
  D(V'\|W|P_n)] \biggr],
\end{align}
which, in turn, shows
\begin{equation}
  \ell(P_n\times V) \doteq \exp\bigl[-n g_n(V\|W|P_n)\bigr]
\end{equation}
with $g_n$ defined as in \eqref{eq:gndef}.  Therefore,
\begin{equation} \label{eq:ccmin}
  \min\{1,\ell(P_n\times V)/M\} \doteq \exp\bigl[-n
  [R-g_n(V\|W|P_n)]^+\bigr].
\end{equation}
Using \eqref{eq:ccmin} in \eqref{eq:ccdivdot} proves \eqref{eq:ccexpn}.
\hfill\IEEEQED
%%%%%%%%%%%%%%%%%%%%%%%%%%%%%%%%%%%%%%%%%%%%%%%%%%%%%%%%%%%%%%%%%%%%%%%%%%%
\section{Conclusion and Discussion}
We studied the \emph{exact} exponential decay rate of the information
leaked to the eavesdropper in Wyner's wiretap channel setting when an
average wiretap channel code in the ensemble of i.i.d.\ or
constant-composition random codes is used for communication.  Our analysis
shows that the previously-derived lower bound on the secrecy exponent of
i.i.d.\ random codes in
\cite{hayashi:2011a,hayashi:2012,han:2014,bastani:2015} is, indeed, tight.
Moreover, our result for constant-composition random codes improves upon
that of \cite{hayashi:2011b} (see \eqref{eq:e0cmp} and examples in
Section~\ref{sec:compare}).

A key step in our analysis (which is applicable to any ensemble of random
codes with independently sampled codewords) is to observe the equivalence
of secrecy and resolvability exponents for the ensemble and, as a result,
reducing the problem to the analysis of the resolvability exponent.
The latter is easier as the informational divergence of interest (whose
exponential decay rate is being assessed) involves a single random
distribution (the output distribution) while the former involves two (the
conditional and unconditional output distributions).   We should emphasize
that establishing secrecy via channel resolvability is a standard technique
which was used in
\cite{csiszar:1996,hayashi:2006,han:2014,bastani:2015,hou:2014} (also, in
  combination with privacy amplification in
\cite{hayashi:2011a,hayashi:2011b}) whose advantages are discussed in
\cite{bloch:2013}.  Our result (Theorem~\ref{thm:exactSecExp}) highlights the
usefulness of this tool by showing that the resolvability exponent is not only
a lower bound to the secrecy exponent but also equals the secrecy exponent.

Thanks to such a reduction, we extended the method of \cite{bastani:2015}
to derive the exact resolvability exponent of random codes.  It is
noteworthy that, as it was already envisioned in \cite{bastani:2015}, the
method presented there was conveniently applicable to the ensemble of
constant-composition random codes (as well as the ensemble of i.i.d.\
random codes already studied in \cite{bastani:2015}). 

It is remarkable that, unlike the channel coding problem for which
constant-composition random codes turn out to be never worse than i.i.d.\
random codes in terms of the exponent \cite{csiszar:it},  for the
secrecy problem we have examples (see Figures~\ref{fig:z} and
\ref{fig:bac}) where i.i.d.\ random codes perform better than
constant-composition codes.  The examples presented in
Section~\ref{sec:compare} suggest that the superior ensemble (in terms of
  the secrecy exponent) depends on the channel $W_\rE$ alone (i.e., for a
  given channel, either of the ensembles yields a better secrecy exponent
for all input distributions). A subject for future research would
be to characterize the set of channels for which the ensemble of i.i.d.\
random codes results in a better secrecy exponent (and vice versa).   

% Constant-composition random codes are particularly instrumental in deriving
% universally achievable exponents \cite{korner:1980}.  
% Combining the results of \cite{korner:1980} and \cite{merhav:2014}, we can
% easily see that, in the wiretap channel setting, when Bob implements a
% maximum mutual information (MMI) decoder, and an average
% constant-composition wiretap channel code is used by Alice, the decoding
% error probability of Bob is exponentially small in the block-length with an
% exponent \emph{equal} to the random coding exponent for
% constant-composition codes (derived in \cite[Chapter 10]{csiszar:it}).
% Note that no knowledge about the channel needs to be available to
% Alice or Bob, thus the exponent is \emph{universally} achievable. It is
% clear that, in such setting, the exact secrecy exponent for random
% constant-composition codes derived in this paper characterizes the
% exponential decay rate of the information Eve learns about the secret
% message (again without requiring any knowledge of channel at the encoder)
% hence is a universally achievable secrecy exponent as defined in
% \cite{hayashi:2011b}.

As shown in \cite{csiszar:1978}, for general pairs of channels
$(W_\rM,W_\rE)$, the secrecy capacity is given by
\begin{equation}
  \max_{\substack{P_{UX}:\\ U\markov X\markov (Y,Z)}} \{I(U;Y)-I(U;Z)\}.
\end{equation}
The secrecy capacity equals 
\begin{equation}
  \max_{P_X}\{I(X;Y)-I(X;Z)\}
\end{equation}
when $\forall P_X$, $I(X;Y)\ge I(X;Z)$.  Accordingly, for the general case
and when the secrecy capacity is positive, one can construct wiretap
channel codes by prefixing the channel with an auxiliary channel
$P_{X|U}:\cU\to\cX$.  Channel prefixing is also
proposed in \cite{han:2014} as a technique to treat the wiretap channels
with cost constraints. (The auxiliary channel $P_{X|U}$ will be chosen such
  that its output sequence satisfies the cost constraints for the physical
  channel.)  It is obvious that our results (as well as those of others
cited) are immediately extensible to such cases.  More precisely,
for a given auxiliary channel $P_{X|U}$, the exponents of \eqref{eq:iidexp}
and \eqref{eq:ccexp}, evaluated for the effective channel $P_{Z|U}(z|u) =
\sum_{x} P_{X|U}(x|u) W_\rE(z|x)$ (instead of $W_\rE$) and the input
distribution $P_U$ are the
ensemble-optimal secrecy exponents of both random-coding ensembles.
Observe that in this setting $P_{X|U}$ (in addition to
the random-binning rate $R$) is also a design parameter which can be
exploited to optimize the secrecy exponent.\footnote{The authors thank the
anonymous reviewer for bringing this point to their attention.}  Moreover,
it should also be noted that in the prefixed setting, in addition to the
entropy rate of $R$ bits per channel use (for random binning), the encoder
requires an entropy rate of $H(X|U)$ bits per channel use to
simulate the channel $P_{X|U}$ that has to be taken into account in
comparison of the secrecy exponents.
%%%%%%%%%%%%%%%%%%%%%%%%%%%%%%%%%%%%%%%%%%%%%%%%%%%%%%%%%%%%%%%%%%%%%%%%%%%
\appendices
\section{Proof of Theorem~\ref{thm:existence}}\label{app:existence}
Consider the sequence of random wiretap channel codes of secret message
size $2 M_\rs$, $M_\rs = \exp(nR_\rs)$ and random binning rate $R$ in the
sense of Definition~\ref{def:wiretapCode}.  Namely, those obtained by
partitioning a random code of size $2\exp[n(R+R_\rs)]$ into $2M_\rs$
sub-codes of rate $R$.  (Assume $R$ and $R_\rs$ are chosen such
$\exp[n(R+R_\rs)]$, $\exp(nR_\rs)$ and $\exp(n R)$ are all integers for
notational brevity.) Let
\begin{align}
  \bar{P}_{\re,n} & \triangleq \E[\Pr\{\hat{s}_{\rm ML}(Y^n) \ne S\}], \\
  \bar{D}_n & \triangleq \E[D(P_{\cC_n^S}\|\bar{P}_{Z^n}|P_S)].
\end{align}
when $S$ is uniformly distributed on $\{1,2,\dotsc,2M_\rs\}$ with
$Y^n$ and $Z^n$ being the output sequences of the legitimate receiver's and
wiretapper's channel respectively as in Figure~\ref{fig:wiretap},
$P_{\cC_n^s}$ being the distribution of wiretapper's channel output
sequence when a uniformly chosen codeword from the sub-code $\cC_n^s$ is
transmitted (see \eqref{eq:output}) and $\bar{P}_{Z^n}$ the distribution
induced by codeword sampling distribution at the output of wiretapper's
channel (see \eqref{eq:ref}). (The
expectation is taken over the choice of codebook
$\cC_n=\bigcup_{s=1}^{2M_\rs} \cC_n^s$)  By the
assumptions of Theorem (in particular, the continuity of $E_\rr$
in rate) and the linearity of expectation we have
\begin{align}
  \liminf_{n\to\infty}-\frac1n\log(\bar{P}_{\re,n})&\ge
  \underline{E}_{\rr}(\Pi,W_\rM,R_\rs+R), \label{eq:existence:peexp}
  \\
  \liminf_{n\to\infty}-\frac1n\log(\bar{D}_n)&\ge
  \underline{E}_{\rs}(\Pi,W_\rE,R). \label{eq:existence:divexp}
\end{align}
Markov's inequality implies that for each $n$, with probability at least
$\frac23$ over the choice of random codes
\begin{equation} \label{eq:existence:pe}
  \Pr\{\hat{s}_{\rm ML}(Y^n)\ne S\} = \frac{1}{2M_\rs}
  \sum_{s=1}^{2M_\rs} \Pr\{\hat{s}_{\rm ML}(Y^n)\ne S| S=s\}
  \le 3 \bar{P}_{\re,n},
\end{equation}
and, with probability at least $\frac23$ 
\begin{equation} \label{eq:existence:div}
  D(P_{\cC_n^S}\|\bar{P}_{Z^n}|P_S) = \frac{1}{2M_\rs}\sum_{s=1}^{2M_\rs}
  D(P_{\cC_n^s}\|\bar{P}_{Z^n}) \le 3 \bar{P}_{\re,n}.
\end{equation}
Therefore, with probability at least $\frac13$, the random code is chosen
such that both bounds of \eqref{eq:existence:pe} and
\eqref{eq:existence:div} simultaneously hold. Let $\cC_n^s$,
$s\in\{1,2,\dotsc,2 M_\rs\}$ be the collection of sub-codes that define any
such good code.  Since the summands in the summation of
\eqref{eq:existence:pe} are all positive, there exists a
subset $\cS_{n,\re} \subseteq \{1,2,\dotsc, 2 M_\rs\}$ of cardinality
$\abs{\cS_{n,\re}} > \frac32 M_\rs$ such that $\forall s \in \cS_{n,\re}$,
\begin{equation} \label{eq:existence:pemax}
  \Pr\{\hat{s}_{\rm ML}(Y^n)\ne S|S=s\} \le 12 \bar{P}_{\re,n}.
\end{equation}
Similarly, since the summands in \eqref{eq:existence:div} are positive,
there exists a subset $\cS_{n,\rs}\subseteq\{1,2,\dotsc,2 M_\rs\}$ of
cardinality $\abs{\cS_{n,\rs}}>\frac32M_\rs$ such that $\forall s \in
\cS_{n,\rs}$ 
\begin{equation}\label{eq:existence:dmax}
  D(P_{\cC^s}\|\bar{P}_{Z^n}) \le 12 \bar{D}_n.
\end{equation}
Pick any $\cS_n \subseteq \cS_{n,\re} \cap \cS_{n,\rs}$ of cardinality
$\abs{\cS_n} = M_\rs$ (this is possible since $\abs{\cS_{n,\re} \cap
\cS_{n,\rs}} \ge M_\rs$) and consider the wiretap channel code that
associates the sub-code $\cC_n^s$ to each message $s\in\cS_n$.  This is a
code of secret message rate $R_\rs$ and, when it is employed with any prior
$P_S$ on secret messages, satisfies 
\begin{equation}\label{eq:existence:pes}
  \Pr\{\hat{s}_{\rm ML}(Y^n)\ne S\} \le 12 \bar{P}_{\re,n},
\end{equation}
due to \eqref{eq:existence:pemax}, and
\begin{equation}\label{eq:existence:infs} 
  I(S;Z^n) \le D(P_{\cC^s_n}\|\bar{P}_{Z^n}|P_S) \le 12 \bar{D}_n,
\end{equation}
due to \eqref{eq:existence:dmax}.  Using this sequence of expurgated codes
we will have
\begin{align} 
  & \liminf_{n\to\infty} -\frac1n \log \Pr\{\hat{s}_{\rm ML}(Y^n)\ne S\} \ge
  \liminf_{n\to\infty} -\frac1n \bar{P}_{\re,n} \nonumber\\
  & \qquad \ge 
  \underline{E}_{\rr}(\Pi,W_\rM,R+R_\rs)
\end{align}
by combining \eqref{eq:existence:pes} and \eqref{eq:existence:peexp}, and
\begin{equation}
  \liminf_{n\to\infty} -\frac1n \log I(S;Z^n) \ge
  \liminf_{n\to\infty} -\frac1n \bar{D}_{n} \ge 
  \underline{E}_{\rs}(\Pi,W_\rE,R)
\end{equation}
by combining \eqref{eq:existence:infs} and \eqref{eq:existence:divexp},
respectively. \hfill\IEEEQED
\begin{remark*}
  The secrecy part of the proof hinges on finding $\exp(nR_\rs)$ ``good''
  resolvability codes via expurgation: we first generated twice as
  many resolvability codes as we needed and then threw away the ``bad''
  half.  Very recently, in \cite{cuff:2016}, it was shown that the
  probability of choosing a bad resolvability code, namely a code $\cC_n$
  (of block-length $n$) for which the $\ell_1$ distance between the output
  distribution $P_{\cC_n}$ \eqref{eq:output} and the reference measure
  $\bar{P}_{Z^n}$ is more than $\exp(-n \gamma)$ for some exponent
  $\gamma$, is \emph{doubly exponentially small in $n$}.  This suggests
  that even if we draw $\exp(n R_\rs)$ codes in a single-shot from the
  ensemble, with very high probability they are \emph{all} good
  resolvability codes.  Nevertheless, we do not know if the results of
  \cite{cuff:2016} hold for the exponents presented in this work.  (Also in
    this work we measure the approximation quality by KL divergence as
    opposed to $\ell_1$ norm but, at least for the i.i.d.\ random coding
    ensemble the KL divergence has the same exponential decay rate as
  the $\ell_1$ distance \cite[Equation~(30)]{cuff:2013}.) 
\end{remark*}
%%%%%%%%%%%%%%%%%%%%%%%%%%%%%%%%%%%%%%%%%%%%%%%%%%%%%%%%%%%%%%%%%%%%%%%%%%%
\section{Proof of Theorem~\ref{thm:asymptotic}}\label{app:lim}
The results when $I(P_X,W)=0$ are trivial. So we only proceed with
the proofs for the case $I(P_X,W)>0$.
\subsection{Proof of (i)}
Let $P_{XZ} = P_X\times W$ for the sake of brevity.  We need to show that
\begin{equation}\label{eq:iidlim:foo}
  \lim_{n\to\infty} E_{\rs,n}^\iid(P_X,W,R) = E_{\rs}^{\iid}(P_X,W,R).
\end{equation}
Recall that $E_{\rs,n}^\iid$ and $E_{\rs}^\iid$ are defined in
\eqref{eq:iidexpn} and \eqref{eq:iidexp} respectively.
Since $\cP_n(\cX\times\cZ)\subset\cP(\cX\times\cZ)$ we trivially have
\begin{equation} \label{eq:iidlim:bar}
  \lim_{n\to\infty} E_{\rs,n}^\iid(P_X,W,R) \ge E_\rs^\iid(P_X,W,R)
\end{equation}
Let $Q^\star$ be the minimizing
distribution in the right-hand-side of \eqref{eq:iidexp}.  Since
$\bigcup_{n\in\NN} \cP_n(\cX\times\cZ)$ is dense in $\cP(\cX\times\cZ)$,
there exists a sequence of $n$-types $\{Q^\star_n \in
\cP_n(\cX\times\cZ)\}_{n\in\NN}$
such that $\lim_{n\to\infty}\abs{Q^\star_n-Q^\star}= 0.$
We, also have,
\begin{equation}\label{eq:iidlim:baz}
  D(Q^\star_n\|P_{XZ})+[R-f(Q^\star_n\|P_{XZ})]^+ \ge E_{\rs,n}^\iid(P_X,W,R)
\end{equation}
Moreover we note that $Q^\star\ll P_{XZ}$ (for if it is not
$D(Q^\star\|P_{XZ})=+\infty$ and $Q^\star$ cannot be the minimizer).
Consequently, we can assume $\forall n\in\NN$, $Q^\star_n \ll P_{XZ}$. 
Since both $D(Q\|P_{XZ})$ and $f(Q\|P_{XZ})$ are continuous in $Q$ over the
set of distributions $Q$ that are absolutely continuous with respect to
$P_{XZ}$,
\begin{align}
  & \lim_{n\to\infty}
  D(Q^\star_n\|P_{XZ})+[R-f(Q^\star_n\|P_{XZ})]^+ \nonumber \\
  & \quad = 
  D(Q^\star\|P_{XZ})+[R-f(Q^\star\|P_{XZ})]^+ \\
  & \quad =  E_{\rs}^\iid(P_X,W,R).
\end{align}
Using \eqref{eq:iidlim:baz} in the above yields,
\begin{equation}
  E_\rs^\iid(P_X,W,R) \ge \lim_{n\to\infty} E_{\rs,n}^\iid(P_X,W,R) 
\end{equation}
which, together with \eqref{eq:iidlim:bar} prove \eqref{eq:iidlim:foo}.
%%%%%%%%%%%%%%%%%%%%%%%%%%%%%%%%%%%%%%%%%%%%%%%%%%%%%%%%%%%%
\subsection{Proof of (ii)}
\subsubsection{Preliminaries}
Let us first examine some properties of the functions $g$ and $g_n$ defined
in \eqref{eq:gdef} and \eqref{eq:gndef} respectively.   To this end, it is
more convenient to look at $g$ and $g_n$ as mappings from the joint
distribution $Q=P\times V\in\cP(\cX\times\cZ)$ to $\RR$, namely,
\begin{align} 
  g(Q,W) & \triangleq % g(Q_{Z|X}\|W|Q_X) = 
  \sum_{x,z} Q(x,z) \log W(z|x) + H(Q_Z) \nonumber\\
  & \quad +
  \min_{\substack{Q'\in\cP(\cX\times\cZ)\colon\\Q'_X=Q_X,\\Q'_Z=Q_Z}}
  D(Q'\|Q'_X\times W), \label{eq:cclim:g} \\
  g_n(Q,W) &\triangleq % g_n(Q_{Z|X}\|W|Q_X) = 
  \sum_{x,z} Q(x,z) \log W(z|x) + H(Q_Z) \nonumber \\
  & +
  \min_{\substack{Q'\in\cP_n(\cX\times\cZ)\colon\\Q'_X=Q_X,Q'_Z=Q_Z}}
  D(Q'\|Q'_X\times W), \label{eq:cclim:gn}
\end{align}
Let us also define the sets $\cQ\subseteq\cP(\cX\times\cZ)$ and
$\cQ_n\subseteq\cP_n(\cX\times\cZ)$ as 
\begin{align} 
  \cQ &\triangleq \{Q\in\cP(\cX\times\cZ)\colon Q \ll Q_X \times W\}.
  \label{eq:cclim:qdef} \\
  \cQ_n &\triangleq \{Q\in\cP_n(\cX\times\cZ)\colon Q\ll Q_X\times W\}. 
  \label{eq:cclim:qndef} 
\end{align}
(Note that $\cQ_n = \cP_n(\cX\times\cZ) \cap \cQ$.)
The set $\cQ$ is compact and convex.
\begin{lemma} \label{lem:gcc}
  The function $g(Q,W)$ defined in \eqref{eq:cclim:g} is continuous in $Q$
  over the set of distributions $Q\in\cQ$.
\end{lemma}
\begin{IEEEproof}
  The linear part $\sum_{x,z} Q(x,z)\log W(z|x)$ is continuous in $Q$ as
  long as $Q(x,z)=0$ whenever $W(z|x)=0$ (which is the case for $Q\in\cQ$).
  The entropy $H(Q_Z)$ is also continuous.  It remains to prove the
  continuity of the last minimization.  We first note that
  \begin{equation}
    \min_{\substack{Q'\in\cP(\cX\times\cZ)\colon\\Q'_X=Q_X,\\Q'_Z=Q_Z}}
    D(Q'\|Q'_X\times W) 
    = 
    \min_{\substack{Q'\in\cQ\colon\\Q'_X=Q_X,\\Q'_Z=Q_Z}}
    D(Q'\|Q'_X\times W)
  \end{equation}
  (for if $Q'\not\in\cQ$, $D(Q'\|Q'_X\times W)=+\infty$ while $Q'=Q$ is a
  feasible point for the minimization where the objective functions has a
  finite value).  The minimum in the above is well-defined as $\cQ$ is
  compact. Let 
  \begin{equation} \label{eq:g:phi}
    \phi(Q)\triangleq
    \min_{\substack{Q'\in\cQ\colon\\Q'_X=Q_X,Q'_Z=Q_Z}}
    D(Q'\|Q'_X\times W).
  \end{equation}
  We prove that $\phi(Q)$ is convex in $Q$:
  Take two distributions $Q_1$ and $Q_2$ in $\cQ$ and let
  $Q=\lambda Q_1 + \overline{\lambda} Q_2$ for some $\lambda\in[0,1]$
  (where we use the short-hand notation of $\overline{\lambda}=1-\lambda$).
  Let 
  \begin{equation}
    Q^\star_j \triangleq
    \argmin_{\substack{Q'\in\cQ:\\Q'_X=(Q_j)_X,Q'_Z=(Q_j)_Z}}
    D(Q'\|Q'_X\times W), \qquad j=1,2,
  \end{equation}
  be the minimizers of \eqref{eq:g:phi}.
  We, hence, have
  \begin{align}
    & \lambda \phi(Q_1) + \overline{\lambda} \phi(Q_2) \nonumber \\
    & \quad = 
    \lambda D(Q_1^\star\|(Q_1^\star)_X\times W) + 
    \overline{\lambda} D(Q_2^\star\|(Q_2^\star)_X \times W)
    \\
    & \quad \age
    D(\lambda Q_1^\star + \overline{\lambda} Q_2^\star \| 
    \lambda (Q_1^\star)_X\times W+\overline{\lambda}(Q_2^\star)_X\times W)
    \\
    & \quad \bge \min_{\substack{Q'\in\cQ:\\
    Q'_X = Q_X, Q'_Z=Q_Z}} D(Q'\|Q'_X\times W) = \phi(Q).
    \label{eq:g:phiconv}
  \end{align}
  where (a) follows since KL divergence is convex in both arguments
  \cite[Lemma~3.5]{csiszar:it}, and
  (b) follows since the joint distribution $\lambda Q_1^\star +
  \overline{\lambda} Q_2^\star$ has $x$-marginal equal to $Q_X$ and
  $z$-marginal equal to $Q_Z$.  The convexity of $\phi$ implies its
  continuity in the interior of the set $\cQ$.  The only discontinuity
  points of $\phi$ could be at the boundaries of the set $\cQ$ where it may
  jump up.  We prove that this cannot happen.

  Let $\{Q_n\in\cQ\}_{n\in\NN}$ be a
  sequence of distributions and $Q = \lim_{n\to\infty} Q_n$ be its
  limit point in $\cQ$. 
  Let 
  \begin{equation} 
    Q^\star_n\triangleq\argmin_{%
      \substack{Q'\in\cQ: \\Q'_X=(Q_n)_X,Q'_Z=(Q_n)_Z}
    }
    D(Q'\|Q'_X\times W)
  \end{equation}
  and $Q^\star = \lim_{n\to\infty} Q_n^\star$ (by passing
  to a subsequence if necessary). 
  Since $D(Q\|Q_X\times W)$ is continuous in $Q$ when $Q\ll Q_X\times W$,
  \begin{equation}
    \lim_{n\to\infty}\phi(Q_n)=D(Q^\star\|Q^\star_X\times W).
  \end{equation}
  Moreover, since $(Q^\star_n)_X=(Q_n)_X$, by continuity of projection we have
  $Q^\star_X = \lim_{n\to\infty} (Q^\star_n)_X = \lim_{n\to\infty}
  (Q_n)_X = Q_X$.  Similarly, $Q^\star_Z=Q_Z$.  Thus,
  \begin{align}
    & \lim_{n\to\infty} \phi(Q_n) = D(Q^\star\|Q^\star_X\times W)
    \nonumber\\
    & \quad \ge
    \min_{\substack{Q'\in\cQ:\\Q'_X=Q_X,\\ Q'_Z=Q_Z}} D(Q'\|Q'_X\times W)
    =
    \phi(Q),
  \end{align}
  which shows $\phi(Q)$ cannot jump up, hence, $\forall Q\in\cQ$, is
  continuous.
\end{IEEEproof}
\begin{remark*}
  It can be checked that for a fixed $P$ and $W$, the function $g(V\|W|P)$, defined
  in \eqref{eq:gdef}, is convex in $V$.
\end{remark*}
\begin{lemma}\label{lem:gncc}
  Let $\{Q_n \in \cQ_n\}_{n\in\NN}$ be a sequence of $n$-types and
  $Q=\lim_{n\to\infty} Q_n\in\cQ$ its limit point (note that since $Q_n \in
    \cQ$ and $\cQ$ is compact, by passing to a subsequence if necessary,
  the limit exists).  Then, 
  \begin{equation} \label{eq:ggn}
    \lim_{n\to\infty} g_n(Q_n,W) = g(Q,W)
  \end{equation}
  (where $g_n(Q_n,W)$ and $g(Q,W)$ are defined in \eqref{eq:cclim:g} and
  \eqref{eq:cclim:gn} respectively).
\end{lemma}
\begin{IEEEproof}
  Same considerations as in the proof of Lemma~\ref{lem:gcc} shows that
  when $Q\in\cQ_n$, the minimizing $Q'$ on the right-hand-side of
  \eqref{eq:cclim:gn} must be in $\cQ_n$.  Define (for $Q\in\cQ_n$),
  \begin{equation} \label{eq:g:phin}
    \phi_n(Q)\triangleq
    \min_{\substack{Q'\in\cQ_n\colon\\Q'_X=Q_X,Q'_Z=Q_Z}}
    D(Q'\|Q'_X\times W).
  \end{equation}
  Since the linear term $\sum_{x,z}Q(x,z)\log W(z|x)$ (for $Q\in\cQ$) and
  entropy $H(Q_Z)$ are continuous, it is sufficient to prove 
  \begin{equation}
    \lim_{n\to\infty} \phi_n(Q_n) = \phi(Q)
  \end{equation}
  where $\phi(Q)$ is defined in \eqref{eq:g:phi}.  Since $\cQ_n\subset\cQ$,
  we trivially have $\phi_n(Q_n)\ge\phi(Q_n)$ and since $\phi$ is
  continuous (as shown in Lemma~\ref{lem:gcc}), we have
  \begin{equation}
    \lim_{n\to\infty} \phi_n(Q_n) \ge \phi(Q).
  \end{equation}
  To prove the reverse inequality, let
  \begin{equation}
    Q^\star \triangleq \argmin_{\substack{Q'\in\cQ:\\Q'_X=Q_X,Q'_Z=Q_Z}}
    D(Q'\|Q'_X\times W).
  \end{equation}
  Since the union of $n$-types is dense in the simplex,  there exists a
  sequence of $n$-types $\{Q^\star_n\}_{n\in\NN}$ such that $\forall
  n\in\NN$, $Q^\star_n \ll Q^\star$ and $\lim_{n\to\infty}
  \abs{Q^\star_n-Q^\star}=0$, therefore
  $\phi(Q)=\lim_{n\to\infty} D(Q^\star_n\|(Q^\star_n)_X\times W)$.
  Moreover, it is easy to verify that $\forall n$, $Q^\star_n\in\cQ_n$.
  Unfortunately, the $x$- and $z$-marginals of $Q^\star_n$ are \emph{not}
  necessarily equal to to $(Q_n)_X$ and $(Q_n)_Z$ respectively.
  Therefore we cannot immediately lower-bound
  $D(Q^\star_n\|(Q^\star_n)_X\times W)$ by $\phi_n(Q_n)$ to conclude the
  proof.   However, since the marginals of $Q^\star_n$ are \emph{close} to
  $(Q_n)_X$ and $(Q_n)_Z$, by perturbing $Q^\star_n$s we can find a second
  sequence of $n$-types, $\{Q^{\star\star}_n\}_{n\in\NN}$ such that
  \begin{enumerate}[(a)] 
    \item $(Q^{\star\star}_n)_X=(Q_n)_X$ and
      $(Q^{\star\star}_n)_Z=(Q_n)_Z$;
    \item $Q^{\star\star}_n\in\cQ_n$; and
    \item $\lim_{n\to\infty}\abs{Q^{\star\star}_n-Q^\star_n} = 0$.
  \end{enumerate}
  Accepting the existence of such a sequence
  $\{Q^{\star\star}_n\}_{n\in\NN}$ we will have
  \begin{align}
    \phi(Q) &= \lim_{n\to\infty} D(Q^\star_n\|(Q^\star_n)_X\times W) \\
    & = \lim_{n\to\infty} D(Q^{\star\star}_n\|(Q^{\star\star}_n)_X\times
    W)\\
    & \ge \lim_{n\to\infty} \phi_n(Q_n)
  \end{align}
  (where the last inequality follows since
    $D(Q^{\star\star}_n\|(Q^{\star\star}_n)_X\times W) \ge \phi_n(Q_n)$ as
    the $x$- and $z$-marginals of $Q^{\star\star}_n$ are equal to $(Q_n)_X$
  and $(Q_n)_Z$ respectively). This will conclude the proof.

  It remains to show the existence of the sequence
  $\{Q^{\star\star}_n\}_{n\in\NN}$.  More precisely, we shall show that
  $\forall \epsilon > 0$,  $\exists n_0(\epsilon)$ such $\forall n >
  n_0$,  we can find $\delta(x,z):\cX\times\cZ\to\RR$ with the following
  properties:
  \begin{enumerate}
    \item $n \delta(x,z) \in \ZZ$;
    \item with 
      \begin{align}
	\delta_X(x) &\triangleq (Q_n)_X(x) - (Q^\star_n)_X(x), & \quad
	\text{and} \\
	\delta_Z(z) &\triangleq (Q_n)_Z(z) - (Q^\star_n)_Z(z),
      \end{align}
      we have
      $\forall x\in\cX$, $\sum_{z\in\cZ} \delta(x,z) = \delta_X(x)$, and
      $\forall z\in\cZ$, $\sum_{x\in\cX} \delta(x,z) = \delta_Z(z)$.
    \item $\forall (x,z) \in \cX\times\cZ$, $\delta(x,z) + Q^\star_n(x,z) \ge
      0$ with equality if $Q^\star_n(x,z)=0$;
    \item $\abs{\delta}\triangleq\sum_{x,z}\abs{\delta(x,z)} \le \epsilon$.
  \end{enumerate}
  (Note that $\delta(x,z)$ also depends on $n$ but we do not show this
  dependence explicitly to keep the notation simple.) If such $\delta$ can
  be found, $Q^{\star\star}_n(x,z)\triangleq Q^\star_n(x,z)+\delta(x,z)$
  will be an $n$-type (due to the first property) whose $x$- and
  $z$-marginals are $(Q_n)_X$ and $(Q_n)_Z$ respectively (due to the second
  property) and is absolutely continuous with respect to $Q^\star_n$ (due
  to the third property) hence is in $\cQ_n$ and is at distance $\epsilon$
  from $Q^{\star}_n$ (due to the fourth property).

  Pick any
  \begin{equation}\label{eq:gncc:eps}
    \gamma < \min\Bigl\{\frac25 \min_{(x,z)\in\supp(Q^\star)} Q^\star(x,z),
    \frac{\epsilon}{2\abs{\cX}\abs{\cZ}}\Bigr\}.
  \end{equation}
  Then, $\exists n_0(\gamma)$ such that for $\forall n > n_0$,
  $\abs{Q^\star_n-Q^\star}\le \gamma/2$ and $\abs{Q_n-Q}\le\gamma/2$.
  Therefore, in particular,
  \begin{equation}
    \abs{(Q_n^\star)_X-Q^\star_X} = 
    \abs{(Q_n^\star)_X-Q_X} \le \gamma/2
  \end{equation}
  and 
  \begin{equation}
    \abs{(Q_n)_X-Q_X}\le\gamma/2
  \end{equation}
  which, together with the triangle inequality imply,
  \begin{equation}
    \abs{(Q^\star_n)_X-(Q_n)_X}\le\gamma.
  \end{equation}
  Similarly, 
  \begin{equation}
    \abs{(Q^\star_n)_Z-(Q_n)_Z}\le\gamma.
  \end{equation}

  Let $G$ be the ``connectivity graph of the joint distribution
  $Q^\star_n$, namely the bipartite graph $G=(\cX,\cZ,\cE)$
  where there is an edge between $x$ and $z$, $(x,z)\in\cE$, iff
  $Q^\star_n(x,z)>0$.
  Suppose $G$ is connected (we discuss what happens if this is not the
  case later).  Then, it certainly has a spanning tree.  Let
  $T=(\cX,\cZ,\cE')$, $\cE'\subseteq\cE$ be one such tree, and pick any
  vertex $v\in\cX\cup\cZ$ as the root.  Suppose the tree has height $H$.
  Let $\cV=\cX\cup\cZ$ be the set of all nodes of $G$ and $\cV_h$ denote
  the set of vertices at height $h$ in the tree.  For every node
  $v\in\cV_h$, let $p(v) \in \cV_{h-1}$ be the parent of $v$ and
  $\cK(v)=\{u \in \cV_{h+1}: (v,u) \in \cE'\}$ be the children of $v$ (with
  $\cK(v) = \emptyset$ for the leaves).  Consider the following algorithm
  to associate a value $\delta_e$ to each edge of the tree: 
  \begin{algorithmic}[1]
    \FOR {$h = H$ \TO $1$}
    \FOR {$v \in \cV_h$}
    \STATE $\delta_e \gets \delta(v) -\sum_{u\in\cK(v)}\delta_{(v,u)}$
    \ENDFOR
    \ENDFOR
  \end{algorithmic}
  where in line 3 we have used the generic notation
  \begin{equation}
    \delta(v) = \begin{cases} 
      \delta_X(x),& \text{if $v\in\cX$},\\ 
      \delta_Z(z), & \text{if $v\in\cZ$}.  
    \end{cases} 
  \end{equation}
  Finally, set 
  \begin{equation}
    \delta(x,z) = \begin{cases}
      \delta_e & \text{if $(x,z)\in\cE'$}\\
      0 & \text{otherwise.}
    \end{cases}
  \end{equation}
  $\delta:\cX\times\cZ\to\RR$, as obtained above, satisfies all the
  desired four properties:
  \begin{enumerate}
    \item is trivial: if $(x,z)$ is not on the tree $n\delta(x,z)=0$,
      otherwise $\delta(x,z)=\delta_e$, $e=(x,z)$ and $\delta_e$ is the
      sum of multiples of $\frac1n$ thus is itself a multiple of $\frac1n$. 
    \item holds by construction except for the root.  Without loss of
      generality suppose the root is a vertex $x_0\in\cX$.  Then, 
      \begin{equation}
	\sum_{x,z} \delta(x,z) = \sum_{z} \delta_Z(z) = 0.
      \end{equation}
      (since $\delta_Z$ is the difference of two distributions). 
      Therefore,
      \begin{align}
	0 & = \sum_{z} \delta(x_0,z) + \sum_{x\ne x_0} \sum_{z}\delta(x,z)
	\\
	& = \sum_{z} \delta(x_0,z) + \sum_{x\ne x_0} \delta_X(x) 
      \end{align}
      which implies
      \begin{equation}
	\sum_{z}\delta(x_0,z) = -\sum_{x\ne x_0} \delta_X(x) =
	\delta_X(x_0)
      \end{equation} 
      again since $\delta_X$ is the difference of two distributions.
  \end{enumerate}
  Moreover by induction on $T$, we can prove that for every edge
  $e\in\cE'$, 
  \begin{equation} \label{eq:gncc:abs2}
    \delta_e \le \sum_{v\in T_{e}} \abs{\delta(v)},
  \end{equation}
  where $T_e$ is the sub-tree rooted at the highest vertex of $e$.
  By extending the sum in \eqref{eq:gncc:abs2} to the entire tree
  and noting that $\sum_{x} \abs{\delta_X(x)} + \sum_{z} \abs{\delta_Z(z)}
  = \abs{(Q^\star_n)_X-(Q_n)_X}+\abs{(Q^\star_n)_Z-(Q_n)_Z}\le2\gamma$, we
  get the following weaker bound: $\forall (x,z)\in\cX\times\cZ$,
  \begin{equation}\label{eq:gncc:abs}
    \abs{\delta(x,z)}\le2\gamma,
  \end{equation}
  which implies the last two properties:
  \begin{enumerate} 
      \setcounter{enumi}{2}
    \item follows since $\delta(x,z)=0$ if $Q_n^\star(x,z)=0$ (as
      $(x,z)\not\in\cE\supset\cE'$) and 
      \begin{align}
	Q_n^\star(x,z) + \delta(x,z) &\ge Q_n^\star(x,z) - 2\gamma \\
	&\ge Q^\star(x,z) - \frac52\gamma\ge0
      \end{align}
      because of \eqref{eq:gncc:eps}.
    \item follows since 
      \begin{equation}
	\abs{\delta} \le 2\abs{\cX}\abs{\cZ}\gamma\le\epsilon
      \end{equation}
      (again because of \eqref{eq:gncc:eps}).
  \end{enumerate}

  \paragraph*{Disconnected $G$}  
  Suppose for some $n$, $G$ is not connected and is rather union of two
  connected components (the proof can be generalized to any finite
  number of components easily). This means that we can partition
  $\cX$ and $\cZ$ into two subsets as $\cX=\cX_1\cup\cX_2$,
  $\cX_1\cap\cX_2=\emptyset$ and $\cZ=\cZ_1\cup\cZ_2$,
  $\cZ_1\cap\cZ_2=\emptyset$ where $\cE = \supp(Q_n^\star)
  \subseteq (\cX_1\times\cZ_1) \cup (\cX_2\times\cZ_2)$.  
  
  This, together with
  the choice of $\gamma$ in \eqref{eq:gncc:eps} implies $\supp(Q^\star)
  \subseteq (\cX_1\times\cZ_1) \cup (\cX_2\times\cZ_2)$ and hence, $\forall
  n$,
  $\supp(Q^\star_n)\subseteq(\cX_1\times\cZ_1)\cup(\cX_2\times\cZ_2)$.
 
  For $\forall n\in\NN$, let
  \begin{equation}
    \lambda_n \triangleq \sum_{(x,z)\in\cX_1\times\cZ_1} Q_n^\star(x,z) = 1-
    \sum_{(x,z)\in\cX_2\times\cZ_2} Q_n^\star(x,z).
  \end{equation}
  Note that $n\lambda_n$ is an integer and by assumption $\lim_{n\to\infty}
  \lambda_n = Q^\star_X(\cX_1) = Q_X(\cX_1) > 0$ (if this is not the case
  we should have started with a smaller $\cX$) thus
  $\lim_{n\to\infty}n\lambda_n=\infty$.   Similarly, we conclude that
  $n(1-\lambda_n)$ is an integer-valued sequence that goes to infinity as
  $n$ grows.

  Let
  \begin{align}
    {Q^\star_n}'(x,z) & \triangleq \frac{Q_n^\star(x,z)}{\lambda_n}
    \ind\{(x,z)\in\cX_1\times\cZ_1\} & \quad \text{and}\\
    {Q^\star_n}''(x,z) & \triangleq
    \frac{Q_n^\star(x,z)}{\overline{\lambda_n}}
    \ind\{(x,z)\in\cX_2\times\cZ_2\},
  \end{align}
  (where we have used the shorthand notation
  $\overline{\lambda_n}=1-\lambda_n$) and observe that 
  \begin{align}
    & D(Q_n^\star\|(Q_n^\star)_X\times W) \nonumber\\
    & \quad = 
    \lambda_n D\bigl({Q^\star_n}'\| ({Q_n^\star}')_X \times W \bigr) +
    \overline{\lambda_n} D\bigl({Q^\star_n}''\| ({Q_n^\star}'')_X \times W \bigr).
    \label{eq:expand}
  \end{align}
  Note that ${Q^\star_n}'$ (resp.\ ${Q^\star_n}''$) is an $n\lambda_n$-type
  (resp.\ $n\overline{\lambda_n}$-type).
  Define also
  \begin{align}
    Q'_n(x,z) & \triangleq \frac{Q_n(x,z)}{\lambda_n} \ind\{(x,z)\in\cX_1\times\cZ_1\} 
    & \quad \text{and}\\
    Q''_n(x,z) & \triangleq \frac{Q_n(x,z)}{\overline{\lambda_n}}
    \ind\{(x,z)\in\cX_2\times\cZ_2\},
  \end{align}
  and note that $Q'_n$ (resp. $Q''_n$) is also an $n\lambda_n$-type 
  (resp.\ an $n\overline{\lambda_n}$-type).  

  Our argument for connected $G$ shows that there
  exists a sequence of $n\lambda_n$-types
  $\{{Q_n^{\star\star}}'\in\cQ_{n\lambda_n}\}_{n\in\NN}$
  such that $\forall n$, $({Q_n^{\star\star}}')_X=(Q_n')_X$,
  $({Q_n^{\star\star}}')_Z=(Q_n')_Z$ and
  $\lim_{n\to\infty}\abs{{Q^{\star\star}_n}'-{Q^\star_n}'}=0$.  Similarly,
  there exists a sequence of $n\overline{\lambda_n}$-types 
  $\{{Q_n^{\star\star}}''\in\cQ_{n\overline{\lambda_n}}\}_{n\in\NN}$
  such that $\forall n$, $({Q_n^{\star\star}}'')_X=(Q_n'')_X$,
  $({Q_n^{\star\star}}'')_Z=(Q_n'')_Z$ and
  $\lim_{n\to\infty}\abs{{Q^{\star\star}_n}''-{Q^\star_n}''}=0$.
  Therefore, 
  \begin{align}
    & D(Q^\star\|Q^\star_X\times W) = \lim_{n\to\infty}
    D(Q^\star_n\|(Q^\star_n)_X\times W) \\ 
    & \quad = \lim_{n\to\infty} \Bigl\{
      \lambda_n D\bigl({Q^\star_n}'\| ({Q_n^\star}')_X \times W \bigr)
      \nonumber\\
      & \quad\qquad +
      \overline{\lambda_n} D\bigl({Q^\star_n}''\| ({Q_n^\star}'')_X \times W
    \bigr)\Bigr\}  \\
    & \quad = \lim_{n\to\infty} \Bigl\{
      \lambda_n D\bigl({Q^{\star\star}_n}'\|({Q_n^{\star\star}}')_X\times
      W\bigr) \nonumber \\ 
      & \quad\qquad +
      \overline{\lambda_n} D\bigl({Q^{\star\star}_n}''\| ({Q_n^{\star\star}}'')_X \times W
    \bigr)\Bigr\}  \\
    & \quad \ge \lim_{n\to\infty} \Bigl\{
      \lambda_n \phi_{n\lambda_n}(Q'_n) + \overline{\lambda_n}
    \phi_{n\overline{\lambda_n}}(Q''_n)\Bigr\}. \label{eq:gncc:foo}
  \end{align}
  Moreover, using the same reasoning as we had to prove convexity of $\phi$
  (see \eqref{eq:g:phiconv}) it follows that
  \begin{equation} 
    \lambda_n \phi_{n\lambda_n}(Q'_n) + \overline{\lambda_n}
    \phi_{n\overline{\lambda_n}}(Q''_n) \ge \phi_n\bigl(\lambda_n Q'_n +
    \overline{\lambda_n} Q''_n\bigr) 
    = \phi(Q_n).
  \end{equation}
  Therefore, continuing \eqref{eq:gncc:foo}, we will again have
  \begin{equation}
    \phi(Q) = D(Q^\star\|Q^\star_X\times W) \ge \lim_{n\to\infty}
    \phi_n(Q_n)
  \end{equation}
  which concludes the proof.
\end{IEEEproof}
\subsubsection{Proof of \eqref{eq:cclim}}
Now we are ready to prove \eqref{eq:cclim}. We need to show that 
\begin{equation}\label{eq:cclim:foo}
  \lim_{n\to\infty} E_{\rs,n}^\cc(P_n,W,R) = E_\rs^\cc(P_X,W,R)
\end{equation}
for any sequence of $n$-types, $P_n\in\cP_n(\cX)$ that converge to $P_X$.
Let 
\begin{equation}
  \tilde{V}_n \triangleq 
  \argmin_{\substack{V\colon\cX\to\cZ:\\P_X \times
  V\in\cP_n(\cX\times\cZ)}} 
  \bigl\{D(V\|W|P_n) + [R - g_n(V\|W|P_n)]^+\bigr\}
\end{equation}
and (by passing to a subsequence if necessary) $\tilde{V} \triangleq
\lim_{n\to\infty} \tilde{V}_n$.  We know that $P_n\times V_n \ll P_n \times
W$, thus, by the continuity of divergence and \eqref{eq:ggn},
\begin{align}
  & \lim_{n\to\infty} E_{\rs,n}^\cc(P_n,W,R) \nonumber \\
  & \quad  = D(\tilde{V}\|W|P_X) + [R-g(\tilde{V}\|W|P_X)]^+ \\ 
  & \quad \ge \min_{V:\cX\to\cZ} \bigl\{ D(V\|W|P_X) + [R-g(V\|W|P_X)]^+  
  \bigr\} \\
  & \quad = E_\rs^\cc(P_X,W,R). \label{eq:cclim:bar}
\end{align}
On the other side, let 
\begin{equation}
  V^\star = \argmin_{V:\cX\to\cZ} \{D(V\|W|P_X) + [R-g(V\|W|P_X)]^+\}.
\end{equation}
There exists a sequence of stochastic matrices $V^\star_n:\cX\to\cZ$ such
that, (a) $P_n\times V^\star_n \in \cP_n(\cX\times\cZ)$, (b)
$\lim_{n\to\infty}\abs{P_n\times V^\star_n-P_X\times V^\star}=0$, and (c)
$\forall n$, $P_n\times V^\star_n \ll P_n \times W$. Accepting this
momentarily, by continuity of $D(V\|W|P)$ and \eqref{eq:ggn}, we have
\begin{align} 
  & E_{\rs}^\cc(P_X,W,R) \nonumber \\ 
  & \quad = \lim_{n\to\infty} \bigl\{
  D\bigl(V^\star_n\|W|P_n\bigr) +
  \bigl[R-g_n\bigl(V^\star_n\|W|P_n\bigr)\bigr]^+\bigr\} \\
  & \quad \ge
  \lim_{n\to\infty} E_{n,\rs}^\cc(P_n,W,R)
\end{align}
which, together with \eqref{eq:cclim:bar} yields \eqref{eq:cclim:foo}.

Existence of such $V_n^\star$s already follows from the algorithm we
presented in the proof of Lemma~\ref{lem:gncc} or more simply from the
following argument:
We assumed (without essential loss of generality) that
$\supp(P_X)=\cX$.  Therefore,
the assumption $\lim_{n\to\infty}\abs{P_n-P_X}=0$, implies $\forall
x\in\cX$, $\lim_{n\to\infty}P_n(x)=P_X(x)> 0$, thus
$\lim_{n\to\infty}nP_n(x)=+\infty$.  Pick $\epsilon > 0$.  Therefore
$\exists n_0(\epsilon)$ such that $\forall n > n_0$,
$\abs{P_X-P_n}\le\epsilon/2$. 
Moreover, for each $x$, $V^\star(\cdot|x)$ is the limit
point of a sequence of $n$-types on $\supp\bigl(V^\star(\cdot|x)\bigr)$.
Therefore, for every $x\in\cX$,
$\exists n_x(\epsilon)$ such that for $\forall n>n_x$, there exists an $nP_n(x)$-type
$V^\star_n(\cdot|x)$ such that $\abs{V^\star(\cdot|x)-V^\star_n(\cdot|x)} \le
\epsilon/2$ and $V^\star_n(\cdot|x) \ll V^\star(\cdot|x)$. Finally, we
observe that $P_n \times V^\star_n$ is a $n$-type and for $n >
\max\bigl\{n_0, \max_{x\in\cX} n_x\bigr\}$,
$\abs{P_n\times V^\star_n-P_X\times V^\star} \le \epsilon$.
%%%%%%%%%%%%%%%%%%%%%%%%%%%%%%%%%%%%%%%%%%%%%%%%%%%%%%%%%%%%
\subsection{Strict Monotonicity of $E_\rs^\iid$ and $E_\rs^\cc$ in $R$}
\label{app:incr}
That $E_\rs^\iid$ is strictly increasing in $R$ for $R > I(P_X,W)$ can be
easily seen through the form of \eqref{eq:iidexpalt}: $E_{\rm s}^\iid$ is the
supremum of affine functions of $R$ thus is convex in $R$.  On the other
side, since $F_0(P_X,W,\lambda)$ is a convex function of $\lambda$ passing
through the origin with slope $I(P_X,W)$, $E_{\rs}^{\iid}(P_X,W,R)$ starts to
increase above $0$ once $R$ exceeds $I(P_X,W)$ which means it will be
strictly increasing for $R>I(P_X,W)$.

We only need to prove the claim for $E_\rs^\cc$. (This proof may
also be used to show $E_\rs^\iid$ is strictly increasing in $R$, replacing
$g$ with $f$.)  
Note that 
\begin{align}
  &E_\rs^\cc(P_X,W,R) = \min \Bigl\{\min_{V: g(V\|W|P_X)\ge R}
  D(V\|W|P_X),\nonumber\\
  &\quad\quad\min_{V: g(V\|W|P_X) \le R} \{D(V\|W|P_X) + R-g(V\|W|P_X)\}\Bigr\}.
\end{align}
We first show that for $R > I(P_X,W)$,
\begin{align}
  & E_\rs^\cc(P_X,W,R) \nonumber\\
  & \quad = \min_{V: g(V\|W|P_X) \le R}
  \{D(V\|W|P_X)
  + R - g(V\|W|P_X)\} \\
  & \quad = R + \min_{V:g(V\|W|P_X)\le R} \{D(V\|W|P_X)-g(V\|W|P_X)\}
  \label{eq:incr:exp}
\end{align}
This follows since for $R>I(P_X,W)$,
\begin{align}
  & \min_{V: g(V\|W|P_X)\ge R} D(V\|W|P_X) \nonumber\\
  & \quad = \min_{V: g(V\|W|P_X) = R} D(V\|W|P_X)
  \label{eq:incr:boundary}
\end{align}
Let us first prove \eqref{eq:incr:boundary}:
Suppose this is not the case, i.e., there exists $V^\star$ with
$g(V^\star\|W|P_X)
> R$ such that $D(V^\star\|W|P_X) \le D(V\|W|P_X)$ for every $V$ with
$g(V\|W|P_X) \ge R$.  We can safely assume that $P_X\times V^\star \ll P_X\times
W$ (otherwise $D(V\|W|P_X)=+\infty$ for all $V$ such that $g(V\|W|P_X) \ge R$
and \eqref{eq:incr:exp} automatically follows). Let
$V_\lambda\triangleq\lambda V^\star+(1-\lambda)W$, for $\lambda \in [0,1]$.
It is easy to check that $\forall \lambda\in[0,1]\colon P_X\times V_\lambda
\ll P_X \times W$, thus the mapping  $\lambda \mapsto g(V_\lambda\|W|P_X)$ is
continuous by the continuity of $g$ (see
Lemma~\ref{lem:gcc}) on the interval $[0,1]$.  We know that $g(V_1\|W|P_X) =
g(V_\star\|W|P_X) > R$ and $g(V_0\|W|P_X) = g(W\|W|P_X) = I(P_X,W) < R$.  Therefore, there exists
$\beta \in (0,1)$ for which $g(V_\beta\|W|P_X) = R$.  On the other side, the
convexity of divergence implies
\begin{align}
  D(V_\beta\|W|P_X) & \le \beta D(V^\star\|W|P_X) + (1-\beta)
  D(W\|W|P_X) \\
  & <  D(V^\star\|W|P_X)
\end{align}
since $\beta<1$.  This contradicts the optimality of $V^\star$. 

Now, we show that $E_\rs^\cc(P_X,W,R') > E_{\rm s}^\cc(P_X,W,R)$ for $R' >
R > I(P_X,W)$.  Let
\begin{equation}
  V^* = \argmin_{V: g(V\|W|P_X)\le R'} \{D(V\|W|P_X)-g(V\|W|P_X)\}.
\end{equation}
If $g(V^*\|W|P_X) \le R$, then
\begin{align}
  & E_\rs^\cc(P_X,W,R') = R' +
  D(V^*\|W|P_X) - g(V^*\|W|P_X) \\
  & \quad = R' + \min_{V: g(V\|W|P_X) \le R} \{D(V\|W|P_X) -
  g(V\|W|P_X)\} \\
  & \quad > R + \min_{V: g(V\|W|P_X) \le R} \{D(V\|W|P_X) - g(V\|W|P_X)\}\\
  & \quad = E_\rs^\cc(P_X,W,R)
\end{align}
which proves the claim.

Otherwise, we have $R < g(V^*\|W|P_X) \le R'$. Consider once again the family of
stochastic matrices defined as $V_\lambda \triangleq \lambda V^* +
(1-\lambda) W$.  We know $P_X\times V^* \ll P_X \times W$ (for if it is
not, $D(V^*\|W|P_X) = +\infty$ and $g(V^*\|W|P_X)=-\infty$ which means the
exponent is infinity which is contradiction since $E_\rs^\cc(P_X,W,R')\le
R'-I(P_X,W)$ by taking $V=W$ in \eqref{eq:incr:exp}).
Using the same reasoning as above, since $g(V_1\|W|P_X) > R$ and
$g(V_0\|W|P_X)=I(P_X,W)<R$ one can find $\beta\in(0,1)$ such that
$g(V_\beta\|W|P_X) = R$ and 
\begin{equation} \label{eq:incr:dbeta}
  D(V_\beta\|W|P_X) \le \beta D(V^\star\|W|P_X).
\end{equation}
Moreover, we know that
\begin{align} 
  & D(V_\beta\|W|P_X) = R + [D(V_\beta\|W|P_X) - g(V_\beta\|W|P_X)] \\
  & \quad \ge R +\min_{V:g(V\|W|P_X)\le
  R}\{D(V\|W|P_X)-g(V\|W|P_X)\} \\ 
  & \quad = E_\rs^\cc(P_X,W,R).
  \label{eq:incr:foo}
\end{align}
One the other side,
\begin{align}
  E_\rs^\cc(P_X,W,R') & = R' + D(V^*\|W|P_X) - g(V^*\|W|P_X) \\
  & \age D(V^*\|W|P_X) \\
  & \bge \frac{1}{\beta} D(V_\beta\|W|P_X) \\
  & \cge \frac{1}{\beta} E_\rs^\cc(P_X,W,R) \\
  & \stackrel{\text{(d)}}{>} E_\rs^\cc(P_X,W,R),
\end{align}
where (a) follows since $g(V^\star\|W|P_X) \le R'$, (b) follows from
\eqref{eq:incr:dbeta} and (c) from \eqref{eq:incr:foo} and finally (d)
holds since $\beta < 1$ and $E_\rs^\cc(P_X,W,R) > 0$.
%%%%%%%%%%%%%%%%%%%%%%%%%%%%%%%%%%%%%%%%%%%%%%%%%%%%%%%%%%%%
\subsection{Alternative form of $E_\rs^\iid$} 
Let $P_{XZ}=P_X\times W$ again.
Using the fact that $\max\{a,0\} = \max_{0 \le \lambda \le 1} \lambda a$, 
\begin{align}
  & \min_Q \left\{ D(Q\|P_{XZ}) + [R-f(Q\|P_{XZ})]^+\right\} \nonumber \\
  &\quad= 
  \min_Q \left\{ D(Q\|P_{XZ}) + \max_{0 \le \lambda \le 1}
  \lambda[R-f(Q\|P_{XZ})]\right\}  \\
  &\quad= 
  \min_Q \max_{0 \le \lambda \le 1}  \left\{ \lambda R + D(Q\|P_{XZ}) -
  \lambda f(Q\|P_{XZ}) \right\}  \\
  &\quad\aeq
  \max_{0 \le \lambda \le 1}  \min_Q \left\{ \lambda R + D(Q\|P_{XZ}) -
  \lambda f(Q\|P_{XZ}) \right\}  \\
  &\quad= 
  \max_{0 \le \lambda \le 1}  \left\{ \lambda R + \min_Q\left\{ D(Q\|P_{XZ}) -
  \lambda f(Q\|P_{XZ}) \right\} \right\} \\
  &\quad\beq 
  \max_{0 \le \lambda \le 1}  \left\{ \lambda R - F_0(P_X,W,\lambda)
  \right\}
\end{align}
where (a) follows since $D(Q\|P_{XZ}) - \lambda f(Q\|P_{XZ})$ is convex in $Q$
(recall that $f$ is linear in $Q$) and (b) since
\begin{align}
  & D(Q\|P_{XZ}) - \lambda f(Q\|P_{XZ}) \nonumber\\ 
  &\quad= \sum_{x,z} Q(x,z) \log \frac{Q(x,z)}{P_{XZ}(x,z)^{1+\lambda}
  P_X(x)^{-\lambda} P_Z(z)^{-\lambda}} \\
  &\quad\stackrel{(\ast)}{\ge} 
  - \log \sum_{x,z} P_{XZ}(x,z)^{1+\lambda} P_X(x)^{-\lambda}
  P_Z(z)^{-\lambda}  \\
  &\quad= F_0(P_X,W,\lambda), 
\end{align}
with equality in $(\ast)$ iff $Q(x,z) \propto
P_{XZ}(x,z)^{1+\lambda}P_X(x)^{-\lambda}P_Z(z)^{-\lambda}$.\hfill\IEEEQED
%%%%%%%%%%%%%%%%%%%%%%%%%%%%%%%%%%%%%%%%%%%%%%%%%%%%%%%%%%%%%%%%%%%%%%%%%%%
\section{Proof of~\eqref{eq:e0cmp}} \label{app:e0cmp}
Taking $V'=V$ in \eqref{eq:gdef}, we have $g(V\|W|P)\le I(P,V)$, thus,
\begin{equation}
  R-g(V\|W|P_X)\ge R-I(P_X,V).
\end{equation}
Therefore,
\begin{align}
  & E_\rs^\cc(P_X,W,R) 
  \nonumber \\
  & \quad = \min_{V} 
  \bigl\{D(V\|W|P_X) + [R-g(V\|W|P_X)]^+\bigr\} \\
  &\quad\ge \min_{V} 
  \bigl\{D(V\|W|P_X) + [R-I(P_X,V)]^+\bigr\} \\
  &\quad\aeq \min_{V} 
  \bigl\{D(V\|W|P_X) + \max_{0\le\lambda\le1} \{\lambda R- \lambda
  I(P_X,V)\}\bigr\} \\
  &\quad\beq \max_{0\le\lambda\le1} \bigl\{\lambda R + \min_{V} 
  \{D(V\|W|P_X) - \lambda I(P_X,V)\}\bigr\} 
  \label{eq:compare:maxmin} 
\end{align}
where (a) follows since $[a]^+=\max_{0\le\lambda\le1}
\lambda a$ and (b) by observing that $D(V\|W|P_X)-\lambda I(P_X,V)$
is convex in $V$ for $\lambda\le1$ (and linear in $\lambda$). The
latter holds since $I(P_X,V) = \min_{Q_Z\in\cP(\cZ)}D(V\|Q_Z|P_X)$,
therefore,
\begin{align}
  &D(V\|W|P_X) - \lambda I(P_X,V)\nonumber\\
  &\,= 
  \max_{Q_Z\in\cP(\cZ)} \{D(V\|W|P_X) - \lambda D(V\|Q_Z|P_X)\} \\
  &\,= \max_{Q_Z} \sum_{x,z} P_X(x)V(z|x)
  \log\frac{V(z|x)^{1-\lambda}}{W(z|x)Q_Z(z)^{-\lambda}}  \\
  &\,= \frac1t \max_{Q_Z} \sum_{x,z}P_X(x)V(z|x)
  \log\frac{V(z|x)}{W(z|x)^tQ_Z(z)^{1-t}}.
  \label{eq:compare:baz}
\end{align} 
where we have defined $t\triangleq\frac1{1-\lambda}$ in the last step.
The objective function inside the $\max$ in \eqref{eq:compare:baz} is
convex in $V$ and since the supremum of convex functions is still
convex, the convexity of $D(V\|W|P_X) - \lambda I(P_X,V)$ in $V$
follows.  It can also be seen that the objective function is concave in
$Q_Z$ for $\lambda > 0$ (i.e. $t>1$).  Using this observation we have
\begin{align}
  & \min_{V} \{ D(V\|W|P_X) - \lambda I(P_X,V) \}  \nonumber\\
  &\quad = \frac1t \min_{V} \max_{Q_Z} \sum_{x,z}P_X(x)V(z|x)
  \log\frac{V(z|x)}{W(z|x)^tQ_Z(z)^{1-t}}
  \\
  &\quad = \frac1t \max_{Q_Z} \min_{V}  \sum_{x,z}P_X(x)V(z|x)
  \log\frac{V(z|x)}{W(z|x)^tQ_Z(z)^{1-t}}
  \\
  &\quad\aeq \max_{Q_Z}\left\{-\frac1t
    \sum_{x}P_X(x)\log\sum_{z}W(z|x)^tQ_Z(z)^{1-t}
  \right\}
  \\ 
  &\quad\bge \max_{Q_Z} \left\{-\frac1t \log \sum_x P_X(x) \sum_z
    W(z|x)^tQ_Z(z)^{1-t}
  \right\}
  \\ 
  &\quad= -\min_{Q_Z} \left\{\frac1t\log 
    \sum_z Q_Z(z)^{1-t} \sum_{x}
    P_X(x)W(z|x)^t
  \right\}
  \label{eq:compare:minq}
\end{align}
where (a) and (b) follow by the concavity of logarithm.  KKT conditions
imply the solution to the minimization of \eqref{eq:compare:minq} is 
\begin{equation}
  Q_Z(z) = c
  \biggl(\sum_{x}P_X(x)W(z|x)^t\biggr)^{1/t}
\end{equation}
with $c^{-1} = \sum_{z}
\left(\sum_{x}P_X(x)W(z|x)^t\right)^{1/t}$.
Plugging this into the objective function of \eqref{eq:compare:minq} and
replacing $t=\frac1{1-\lambda}$, we have
\begin{align}
  & \min_{V} \{ D(V\|W|P_X) - \lambda I(P_X,V) \}\nonumber\\
  &\quad= - \log \sum_{z}
  \biggl(\sum_{x}P_X(x)W(z|x)^\frac{1}{1-\lambda}\biggr)^{1-\lambda} \\
  &\quad= -E_0(P_X,W,\lambda). \label{eq:compare:E0}
\end{align} 
Plugging \eqref{eq:compare:E0} into \eqref{eq:compare:maxmin} proves the
claim. \hfill \IEEEQED
%%%%%%%%%%%%%%%%%%%%%%%%%%%%%%%%%%%%%%%%%%%%%%%%%%%%%%%%%%%%%%%%%%%%%%%%%%%
\section{Numerical Evaluation of The Secrecy Exponents} \label{app:compute}
\subsection{Computing $E_\rs^\iid$ and $\underline{E}_{\rm
s}^\cc$}
Both $E_\rs^\iid$ and $\underline{E}_\rs^\cc$ can be easily
evaluated via the expressions \eqref{eq:iidexpalt} and \eqref{eq:ccweak} using
the fact that both $F_0$ and $E_0$ (defined in
\eqref{eq:f0} and \eqref{eq:e0} respectively) are convex in $\lambda$, and
pass through the origin with slope $I(P_X,W)$.

For instance to evaluate $E_\rs^\iid$ we know that
\begin{enumerate}
  \item for $R \le I(P_X,W) = \frac{\partial}{\partial \lambda}
    F_0(P_X,W,\lambda) \bigl|_{\lambda=0}$, $E_\rs(P_X,W,R) = 0$; 
  \item for $I(P_X,W) < R < \frac{\partial}{\partial \lambda}
    F_0(P_X,W,\lambda) \bigl|_{\lambda=1}$, the pairs $R$, $E_{\rm
    s}^\iid$ are related parametrically as
    \begin{subequations}
      \begin{align}
	R(\lambda) &= \frac{\partial}{\partial \lambda} F_0(P_X,W,\lambda)
	\\ 
	E_\rs(\lambda) &= \lambda R(\lambda) - F_0(P_X,W,\lambda)
      \end{align}
    \end{subequations}
    for the range of $\lambda \in [0,1]$; 
  \item finally, if $R \ge F_0'(1)$, 
    \begin{equation}
      E_\rs(P_X,W,R) = R - F_0(P_X,W,1).
    \end{equation}
\end{enumerate}

It is clear that to evaluate $\underline{E}_\rs^\cc$, one has to follow
precisely the same steps replacing $F_0$ with $E_0$. 
%%%%%%%%%%%%%%%%%%%%%%%%%%%%%%%%%%%%%%%%%%%%%%%%%%%%%%%%%%%%%%%%%%%%%%%%%%%
\subsection{Computing $E_\rs^\cc$}
To compute $E_\rs^\cc$ (defined in \eqref{eq:ccexp}), one has
to solve two minimizations.  Namely, that of \eqref{eq:ccexpdef} and
that of \eqref{eq:gdef}.   The latter turns out to be
efficiently solvable using standard convex optimization tools. 

Fix $Q_Z\in\cP(\cZ)$ (to be set to $P_X\circ V$ to compute $g(V\|W|P_X)$).  
We have:
\begin{align}
  &\min_{V': P_X\circ V' = Q_Z} D(V'\|W|P_X)
  = 
  \min_{V'} \Bigl\{ 
    D(V'\|W|P_X) \nonumber\\
    &\quad\qquad+ \max_{\rho \in \RR^{\abs{\cZ}}} 
    \sum_{z} \rho_z \left[Q_Z(z) - (\!P_X\circ V'\!)(z)\right]\Bigr\} \\
  &\quad= 
  \max_{\rho \in \RR^{\abs{\cZ}}} \Bigl\{ \min_{V'} \Bigl\{ D(V'\|W|P_X) -
    \sum_{x,z} P_X(x) V'(z|x) \rho_z \Bigr\} \nonumber\\ 
  &\quad\qquad+\sum_{z} \rho_z Q_Z(z) \Bigr\},
  \label{eq:compute:foo}
\end{align}
where $\rho\triangleq(\rho_1,\dotsc,\rho_{\abs{\cZ}})$ and the last equality
follows since $D(V\|W|P_X)$ is convex in $V$ and the second term is linear
in $V$.  Moreover, the inner unconstrained minimization has the value
\begin{align} 
  &\min_{V'} \Bigl\{D(V'\|W|P_X) - \sum_{x,z} P_X(x) V'(z|x) \rho_z
  \Bigr\} \nonumber\\
  &\quad= 
  \min_{V'}
  \sum_{x,z} P_X(x) V'(z|x) \log \frac{V'(z|x)}{W(z|x) \exp(\rho_z)} 
  \\
  &\quad=
  - \sum_{x} P_X(x) \log \sum_{z} W(z|x) \exp(\rho_z),
\end{align}
by choosing $V'(z|x) \propto W(z|x) \exp(\rho_z)$.  Plugging this into
\eqref{eq:compute:foo}, we get 
\begin{align}
  & \min_{V': P_X\circ V'=Q} D(V'\|W|P_X) 
  = 
  \max_{\rho\in\RR^{\abs{\cZ}}} \Bigl\{ \sum_{z}\rho_z Q_Z(z) \nonumber\\
  &\quad\qquad- \sum_{x}
  P_X(x) \log\sum_{z} W(z|x)\exp(\rho_z)\Bigr\}.
  \label{eq:compute:bar}
\end{align}
\begin{remark*} 
  Using H\"older's inequality, it can be checked that the objective
  function of \eqref{eq:compute:bar} is concave in $\rho$, thus can be
  efficiently maximized using standard numerical methods.
  \begin{IEEEproof}
    Since the first sum in the objective function of \eqref{eq:compute:bar} is
    linear in $\rho$ it is sufficient to prove that the function
    \begin{equation}
      \rho \mapsto \sum_{x} P_X(x) \log\left(W(z|x)\exp(\rho_z)\right)
    \end{equation}
    is convex in $\rho$.  Fix $t \in [0,1]$ and $\rho, \rho' \in
    \RR^{\abs{\cZ}}$.  For every $x\in\cX$, H\"older's inequality implies
    \begin{align}
      &\sum_{z} W(z|x) \exp(t\rho_z + (1-t) \rho'_z) \nonumber\\
      &\quad= \sum_{z}
      W(z|x)^t \exp(t\rho_z) \cdot W(z|x)^{1-t} \exp((1-t)\rho'_z)
      \\
      &\quad\le 
      \left(\sum_{z}W(z|x)\exp(\rho_z)\right)^t \cdot
      \left(\sum_{x}W(z|x)\exp(\rho'_z)\right)^{1-t}
    \end{align}
    Taking the logarithm of both sides, multiplying by $P_X(x)$, and
    finally summing over $x$ proves the claim.
  \end{IEEEproof}
\end{remark*}
Finally, for small alphabet sizes that we have considered in
Section~\ref{sec:compare}, we can solve the minimization of
\eqref{eq:ccexpdef} via exhaustive search. 
%%%%%%%%%%%%%%%%%%%%%%%%%%%%%%%%%%%%%%%%%%%%%%%%%%%%%%%%%%%%%%%%%%%%%%%%%%%
\section{Proof of Lemma~\ref{lem:ref}} \label{app:ref}
\begin{enumerate}[(i)]
  \item
    The linearity of expectation shows that $\bar{P}_{Z^n}$ as defined in
    \eqref{eq:ref} is the expectation of the non-negative random variable
    $P_{\cC_n}(z^n)$ (defined in \eqref{eq:output}).  Therefore,
    $\bar{P}_{Z^n}(z^n)=0$ implies $P_{\cC_n}(z^n)=0$ almost surely.  
  \item
    Pick $z^n$ and $\tilde{z}^n$ that have the same type.  Therefore, there
    exists a permutation, call it $\pi\colon\cZ^n\to\cZ^n$, such that
    $\tilde{z}^n = \pi(z^n)$ and $z^n = \pi^{-1}(\tilde{z}^n)$.  Then,
    \begin{align}
      \bar{P}_{Z^n}(\tilde{z}^n) &= \sum_{x^n} P_{X^n}(x^n)
      W^n(\tilde{z}^n|x^n) \\
      & \aeq \sum_{\tilde{x}^n} P_{X^n}\bigl(\pi(\tilde{x}^n)\bigr)
      W^n\bigl(\pi(z^n)|\pi(\tilde{x}^n)\bigr) \\
      & \beq \sum_{\tilde{x}^n} P_{X^n}(\tilde{x}^n) W^n(z^n|\tilde{x}^n) =
      \bar{P}_{Z^n}(z^n).
    \end{align}
    where in (a) we have taken $x^n=\pi(\tilde{x}^n)$ and (b) follows since
    $P_{X^n}(x^n)$ only depends on the type of $x^n$ (and by
    construction $\tilde{x}^n$ and $\pi(\tilde{x}^n)$ have the same type)
    and similarly
    $W^n\bigl(\pi(z^n)|\pi(\tilde{x}^n)\bigr)=W^n(z^n|\tilde{x}^n)$.
  \item
    We have
    \begin{equation}
      \bar{P}_{Z^n}(z^n) = \sum_{x^n\in\cX^n} P_{X^n}(x^n)W^n(z^n|x^n)
      \label{eq:ref:sum} 
    \end{equation}
    $\bar{P}_{Z^n}(z^n)>0$ implies there exists at least one sequence
    $x^n_0\in\supp(P_{X^n})$ for which $W^n(z^n|x^n_0)>0$.  Therefore,
    $W^n(z^n|x^n_0)>W_{\min}^n$.  Thus \eqref{eq:ref:sum} yields
    \begin{equation} \label{eq:ref:foo}
      \bar{P}_{Z^n}(z^n) \ge P_{X^n}(x^n_0) W_{\min}^n.
    \end{equation}
    For i.i.d.\ random coding ensemble, $P_{X^n}(x^n) = P_X^n(x^n) \ge
    P_{\min}^n$ and for the constant-composition random coding ensemble,
    $P_{X^n}(x^n) = 1/\bigabs{\cT_{P_X}^n}\ge(1/\abs{\cX})^n$ (since
    $\cT_{P_X}^n\subseteq\cX^n$).
    \hfill\IEEEQED
\end{enumerate}
%%%%%%%%%%%%%%%%%%%%%%%%%%%%%%%%%%%%%%%%%%%%%%%%%%%%%%%%%%%%%%%%%%%%%%%%%%%
\section{Proof of Lemma~\ref{lem:alna}} \label{app:alna}
Take $U \triangleq \frac{A}{\E[A]}$ so that $\E[U] = 1$. We shall prove
that 
\begin{equation}
  c(\theta) \left(\var(U) - \tau_\theta(U)\right)
  \le
  \E[U \ln(U)] \le \var(U).
  \label{eq:alna:ulnu}
\end{equation}
The claim then follows by noting that $\E[A \ln(A/\E[A])] =
\E[A] \, \E[U \ln(U)]$ and $\var(A) = \var(U) / (\E[A])^2$.

We first have 
\begin{align}
  \E[U \ln(U)] &= \E[U\ln(U) - (U-1)] \\ &\le \E[(U-1)^2] = \var(U),
\end{align}
since $u \ln(u) - (u-1) \le (u-1)^2$. On the other hand,
\begin{equation}
  u \ln(u) - (u-1) \ge c(\theta) (u-1)^2 \ind\{u \le \theta+1\}.
\end{equation}
This follows by observing that $\frac{u \ln(u) - (u-1)}{(u-1)^2}$ is a
decreasing function of $u$ (see Lemma~\ref{lem:incdec} below).  Thus,
\begin{equation}
  \E[U \ln(U)] \ge c(\theta) \int_0^{\theta+1} (u-1)^2 \rd F_U(u).
\end{equation}
where $F_U(u)$ is the cumulative distribution function of $u$.

Furthermore,
\begin{equation} \label{eq:alna:partial}
  \int_{0}^{\theta+1} (u-1)^2 \rd F_U(u) = \var(U) -
  \int_{\theta+1}^{+\infty} (u-1)^2 \rd F_U(u)
\end{equation}

Let $v \triangleq u-1$ for the sake of brevity and denote by
$\bar{F}_V(v) \triangleq \Pr\{V > v\} = \Pr\{U > v+1\}$ the
complementary distribution function of $V$. Then,
\begin{align}
  & \int_{\theta+1}^{+\infty} (u-1)^2 \rd F_U(u) = 
  \int_{\theta}^{+\infty} v^2 \rd F_V(v) \\
  &\quad= \left[-v^2 \bar{F}_V(v) \right]_{\theta}^{+\infty} + 
  2 \int_{\theta}^{+\infty} v \bar{F}_V(v) \rd v \\
  &\quad\stackrel{(\ast)}{=}
  \theta^2 \bar{F}_V(\theta) + 2 \int_{\theta}^{+\infty} v
  \bar{F}_V(v) \rd v.
  \label{eq:alna:excess}
\end{align}
The equality in ($\ast$) follows since we assumed the variance of $U$
exists.  This proves \eqref{eq:alna:ulnu}.
\hfill\IEEEQEDhere 
\begin{lemma}\label{lem:incdec}
  For $t\ge0$,
  \begin{enumerate}[(i)] 
    \item the mapping $t \mapsto \frac{t \ln(t) - (t-1)}{t-1}$ is
      increasing in $t$;
    \item the mapping $t \mapsto \frac{t \ln(t) - (t-1)}{(t-1)^2}$ is
      decreasing in $t$.
  \end{enumerate}
\end{lemma}
\begin{IEEEproof}\
  \begin{enumerate}[(i)]
    \item\
      \begin{equation}
	\frac{\partial}{\partial t} \Bigl\{ \frac{t \ln(t) - (t-1)}{t-1}
	\Bigr\} = \frac{(t-1) - \ln(t)}{(t-1)^2} \ge 0
      \end{equation}
      since $\ln(t) \le t-1$.
    \item\
      \begin{equation}\label{eq:incdec:foo}
	\frac{\partial}{\partial t} \Bigl\{ \frac{t \ln(t) - (t-1)}{(t-1)^2}
	\Bigr\} = \frac{2(t-1) - (t+1) \ln(t)}{(t-1)^3} \le 0,
      \end{equation}
      since for $t\ge1$, $\ln(t) \ge 2 \frac{t-1}{t+1}$ while for $t\le 1$,
      $\ln(t)\le 2 \frac{t-1}{t+1}$.  The latter follows since $\ln(t) -
      2 \frac{t-1}{t+1}$ equals $0$ at $t=1$ and has derivative
      \begin{equation}
	\frac{(t-1)^2}{t(t+1)^2} \ge 0. \tag*{\IEEEQEDhere}
      \end{equation}
  \end{enumerate}
\end{IEEEproof}
%%%%%%%%%%%%%%%%%%%%%%%%%%%%%%%%%%%%%%%%%%%%%%%%%%%%%%%%%%%%%%%%%%%%%%%%%%%
\section{Proof of \eqref{eq:p}} \label{app:p}
We have
\begin{align}
  p_Q(z^n) & = \sum_{x^n\in\cX^n} \ind\{(x^n,z^n)\in \cT^n_Q\}
  P_{X^n}(x^n) \\
  & = 
  \frac{P_{X^n}(\cT_{Q_X}^n)}{\abs{\cT_{Q_X}^n}}
  \sum_{x^n\in\cX^n} \ind\{(x^n,z^n)\in \cT^n_Q\}
  \label{eq:p:foo}
\end{align}
since $P_{X^n}(x^n)$ only depends on the type of $x^n$.  On the other side,
we have
\begin{equation}
  \label{eq:p:bar}
  \abs{\cT_Q^n} = \sum_{z^n\in\cZ^n} \sum_{x^n\in\cX^n}
  \ind\bigl\{(x^n,z^n) \in \cT_Q^n\bigr\}
\end{equation}
The value of the inner sum in \eqref{eq:p:bar} only depends on
the type of $z^n$ (this can be easily checked using the same type of
argument as we had in Appendix~\ref{app:ref} part (ii)) and, clearly, is
zero if $Q_Z \ne \type{z^n}$.  Thus 
\begin{equation} \label{eq:p:baz}
  \abs{\cT^n_Q} = \abs{\cT^n_{Q_Z}} \ind\{Q_Z = \type{z^n}\}
  \sum_{x^n\in\cX^n} \ind\bigl\{(x^n,z^n)\in\cT^n_Q\bigr\}.
\end{equation}
Plugging \eqref{eq:p:baz} into \eqref{eq:p:foo} yields \eqref{eq:p}.
\hfill\IEEEQED
%%%%%%%%%%%%%%%%%%%%%%%%%%%%%%%%%%%%%%%%%%%%%%%%%%%%%%%%%%%%%%%%%%%%%%%%%%%
\section{Proof of \eqref{eq:stats}} \label{app:sumstat}
We only prove \eqref{eq:nuval} (as \eqref{eq:muval} is trivial).  (We omit
the dependence on $z^n$ throughout the proof for notational brevity.)
\begin{align}
  & \var(L_1) = \sum_{Q \in \cQ'_n} \frac1{M^2}
  \ell(Q)^2 \var(N_Q) \nonumber\\
  & \qquad +
  \sum_{\substack{(Q_1,Q_2) \in {\cQ'_n}^2 \\
  Q_1 \ne Q_2}} 
  \frac1{M^2} \ell(Q_1) \ell(Q_2)
  \cov(N_{Q_1},N_{Q_2}) 
  %\nonumber 
  \\
  & \quad \stackrel{(\star)}= \frac1M \sum_{Q \in \cQ'_n}
  \ell(Q)^2 p_Q (1-p_Q)  \nonumber\\
  & \qquad - \frac1M \sum_{\substack{(Q_1,Q_2)\in{\cQ'_n}^2 \\
  Q_1\ne Q_2}} \ell(Q_1)\ell(Q_2)
  p_{Q_1} p_{Q_2}, \label{eq:varsum:expand}
\end{align}
where $(\star)$ follows since $\var(N_Q) = M p_Q (1-p_Q)$ and
$\cov(N_{Q_1},N_{Q_2}) = - M p_{Q_1} p_{Q_2}$. Moreover,
\begin{align} 
  & \sum_{\substack{(Q_1,Q_2)\in{\cQ'_n}^2 \\
  Q_1\ne Q_2}} \ell(Q_1) \ell(Q_2)
  p_{Q_1} p_{Q_2} 
  \nonumber\\ 
  &\quad= \sum_{Q_1\in\cQ_n'} \ell(Q_1) p_{Q_1}
  \sum_{Q_2\in\cQ_n'\setminus\{Q_1\}}
  \ell(Q_2) p_{Q_2} 
  %\nonumber 
  \\
  &\quad= \sum_{Q_1 \in \cQ_n'} \ell(Q_1) p_{Q_1}
  \Bigl(\E[L_1] - p_{Q_1}\ell(Q_1)\Bigr).
\end{align}
Using the above in \eqref{eq:varsum:expand} we get, 
\begin{align}
  & \var(L_1) \nonumber \\
  &\quad= \frac1M \sum_{Q\in\cQ'_n} \ell(Q)
  p_Q \Bigl[ (1-p_Q) \ell(Q)- \bigl(\E[L_1] -
  p_Q \ell(Q) \bigr) \Bigr] 
  %\nonumber 
  \\
  &\quad=  \frac1M \sum_{Q \in \cQ'_n} 
  \ell(Q) p_Q \bigl[\ell(Q) - \E[L_1]\bigr]
  %\nonumber 
  \\ 
  &\quad= \frac1M \sum_{Q \in \cQ'_n} \ell(Q)^2
  p_Q - \frac1M \E[L_1]^2.
  \tag*{\IEEEQED}
\end{align}
%%%%%%%%%%%%%%%%%%%%%%%%%%%%%%%%%%%%%%%%%%%%%%%%%%%%%%%%%%%%%%%%%%%%%%%%%%%
\section{Proof of \eqref{eq:divdot}} \label{app:divdot}
Equation~\eqref{eq:divdot1} immediately implies
\begin{align}
  \E[D(P_{\cC_n}\|\bar{P}_{Z^n})]
  & \dotle 
  \sum_{Q\in\cP_n(\cX\times\cZ)} \exp\bigl\{-nD(Q\|Q_X\times W)\bigr\}
  \nonumber \\ & \qquad \times P_{X^n}\bigl(\cT_{Q_X}^n\bigr)
  \min\Bigl\{1,\frac{\ell(Q)}{M}\Bigr\}.
  \label{eq:divdot:le}
\end{align}
It remains to show
\begin{align}
  \E[D(P_{\cC_n}\|\bar{P}_{Z^n})]
  & \dotge 
  \sum_{Q\in\cP_n(\cX\times\cZ)} \exp\bigl\{-nD(Q\|Q_X\times W)\bigr\}
  \nonumber \\ & \qquad \times P_{X^n}\bigl(\cT_{Q_X}^n\bigr)
  \min\Bigl\{1,\frac{\ell(Q)}{M}\Bigr\},
  \label{eq:divdot:ge}
\end{align}
to establish \eqref{eq:divdot}.

Equation~\eqref{eq:divdot1} means there exists a sub-exponentially
increasing sequence $\beta(n)$ (which depends only on
$\abs{\cX}$ and $\abs{\cZ}$) such that 
\begin{align}
  & \beta(n) \Big[\E[D(P_{\cC_n}\|\bar{P}_{Z^n})] + \frac{\log(\e)}{M}\Bigr]
  \nonumber\\
  & \quad \ge 
  \sum_{Q\in\cP_n(\cX\times\cZ)} \exp\bigl\{-nD(Q\|Q_X\times W)\bigr\}
  \nonumber \\
  & \quad \phantom{\doteq\sum_{Q\in\cP_n(\cX\times\cZ)}}
  \times P_{X^n}\bigl(\cT_{Q_X}^n\bigr) 
  \min\Bigl\{1,\frac{\ell(Q)}{M}\Bigr\}.
  \label{eq:divdot:ge1}
\end{align}

Since the union of $n$-types is dense in $\cP(\cX\times\cZ)$, for large
enough $n$, there exists an $n$-type that is as close as desired to the
joint distribution $P_X\times W$. More precisely, for every $\epsilon>0$,
there exists $n_0(\epsilon)$ such that $\forall n > n_0(\epsilon)$, there
exists $Q_n\in\cP_n(\cX\times\cZ)$ for which $I(Q_n) \ge
I(P_X,W)-\epsilon$, $D(Q_n\|(Q_n)_X\times W)\le\epsilon/2$ and
$P_{X^n}\bigl(\cT_{(Q_n)_X}^n\bigr)>\exp(-n\epsilon/2)$. Indeed, taking $Q_n
= P_n \times W_n$, where $P_n$ is an $n$-type quantization of $P_X$ for the
i.i.d.\ random coding ensemble and $W_n$ is the quantization of $W$ such
that $W_n(\cdot|x)$ is a $n P_n(x)$-type yields all desired properties.

Note also that
\begin{align}
  & \ell(Q) \ge \exp\bigl(n\omega(Q)\bigr)\bigabs{\cT_{Q_Z}^n} \\
  &\quad \stackrel{(\ast)}{\ge}
  (n+1)^{-\abs{\cZ}} \exp(n [\omega(Q)+H(Q_Z)]) \\
  &\quad = 
  (n+1)^{-\abs{\cZ}} 
  \exp\bigl(n [I(Q)-D(Q\|Q_X\times W)]\bigr),
\end{align}
where $(\ast)$ follows from \cite[Lemma~2.3]{csiszar:it}.
Let 
\begin{equation} \label{eq:divdot:eps}
  \epsilon \triangleq \min \{R/2, I(P_X,W)/4\} >  0
\end{equation} 
and observe that for all $n \ge n_0(\epsilon)$ with $Q_n$ as described
above
\begin{equation}
  \ell(Q_n) \ge (n+1)^{-\abs{\cZ}} \exp\{n
  (I(P_X,W)-2\epsilon)\}.
\end{equation}
Consequently, the term corresponding to $Q=Q_n$ in the summation of
\eqref{eq:divdot:ge1} is lower-bounded as
\begin{align}
  &\exp\bigl(-n D(W_n\|W|P_n)\bigr)
  P_{X^n}\bigl(\cT_{P_n}^n\bigr)
  \min\Bigl\{1,\frac{\ell(Q_n)}{M}\Bigr\}\nonumber \\
  & \quad \ge 
  (n+1)^{-\abs{\cZ}} 
  \exp\{-n (\epsilon + [R-I(P_X,W)+2\epsilon]^+)\}
  \\
  & \quad \ge 
  (n+1)^{-\abs{\cZ}} \exp\{-n (R-\epsilon)\}.  \label{eq:divdot:foo}
\end{align}
The last inequality follows because of the choice of $\epsilon$ in
\eqref{eq:divdot:eps}.
% The last inequality follows since
% \begin{equation}
%   [R-I(P_X,W)+2\epsilon]^+ \le R-2\epsilon
% \end{equation}
% because $2\epsilon\le R$ and $\epsilon\le
% I(P_X,W)/4$ (hence $R-I(P_X,W)+2\epsilon\le R-2\epsilon$).
Obviously, $\exists n_1(\epsilon,|\cX|,|\cZ|)$ such that $\forall n \ge
n_1$,
\begin{align}
  \beta(n) \frac{\log(\e)}{M} &= \beta(n) \log(\e) \exp(-nR) \nonumber\\
  & \le \frac12 (n+1)^{-\abs{\cZ}} \exp\bigl(-n (R-\epsilon) \bigr).
\end{align}
This, together with \eqref{eq:divdot:foo} implies for $n \ge n_2 \triangleq
\max\{n_0,n_1\}$,
\begin{align}
  \beta(n) \frac{\log(\e)}{M} & \le \frac12 
  \exp\bigl(-n D(W_n\|W|P_n)\bigr)
  P_{X^n}\bigl(T_{P_n}^n\bigr) \nonumber \\
  & \qquad\times \min\Bigl\{1,\frac{\ell(Q_n)}{M}\Bigr\}.
  \label{eq:divdot:bar}
\end{align}
Using \eqref{eq:divdot:bar} in \eqref{eq:divdot:ge} (and multiplying the
summands corresponding to $Q\ne Q_n$ by $\frac12$) we conclude that for
$n \ge n_2$,
\begin{align}
  &\beta(n) \E[D(P_{\cC_n}\|\bar{P}_{Z^n})]\nonumber\\
  & \quad \ge 
  \frac12 \sum_{Q\in\cP_n(\cX\times\cZ)} \exp\bigl\{-nD(Q_{Z|X}\|W|Q_X)\bigr\}
  \nonumber \\
  & \quad \phantom{\doteq\sum_{Q\in\cP_n(\cX\times\cZ)}}
  \times P_{X^n}\bigl(\cT_{Q_X}^n\bigr) 
  \min\Bigl\{1,\frac{\ell(Q)}{M}\Bigr\}.
  \label{eq:divdot:ge2}
\end{align}
Take 
\begin{equation}
 \beta'(n) \triangleq \begin{cases}
    +\infty & \text{if $n<n_2$} \\
    2 \beta(n) & \text{otherwise}.
  \end{cases}
\end{equation}
Therefore, $\forall n$,
\begin{align}
  &
  \beta'(n) \E[D(P_{\cC_n}\|\bar{P}_{Z^n})]\nonumber\\
  & \quad \ge 
  \sum_{Q\in\cP_n(\cX\times\cZ)} \exp\bigl\{-nD(Q_{Z|X}\|W|Q_X)\bigr\}
  \nonumber \\
  & \quad \phantom{\doteq\sum_{Q\in\cP_n(\cX\times\cZ)}}
  \times P_{X^n}\bigl(\cT_{Q_X}^n\bigr) 
  \min\Bigl\{1,\frac{\ell(Q)}{M}\Bigr\}.
  \label{eq:divdot:ge3}
\end{align}
We finally have 
\begin{equation}
  \limsup_{n\to\infty} \frac1n \log \beta'(n) = \limsup_{n\to\infty}
  \frac1n \log \beta(n) = 0
\end{equation}
by assumption and that $\beta'$ only depends on $\abs{\cX}$, $\abs{\cZ}$,
$R$, $P_X$, and $W$ (because $n_2$ only depends on these parameters).
Therefore, \eqref{eq:divdot:ge3} establishes \eqref{eq:divdot:ge} and
concludes the proof.
\hfill\IEEEQED
%%%%%%%%%%%%%%%%%%%%%%%%%%%%%%%%%%%%%%%%%%%%%%%%%%%%%%%%%%%%%%%%%%%%%%%%%%%
% % use section* for acknowledgment
% \section*{Acknowledgment}
% 
% 
% The authors would like to thank...

% Can use something like this to put references on a page
% by themselves when using endfloat and the captionsoff option.
\ifCLASSOPTIONcaptionsoff
  \newpage
\fi

% trigger a \newpage just before the given reference
% number - used to balance the columns on the last page
% adjust value as needed - may need to be readjusted if
% the document is modified later
%\IEEEtriggeratref{8}
% The "triggered" command can be changed if desired:
%\IEEEtriggercmd{\enlargethispage{-5in}}

% references section

% can use a bibliography generated by BibTeX as a .bbl file
% BibTeX documentation can be easily obtained at:
% http://mirror.ctan.org/biblio/bibtex/contrib/doc/
% The IEEEtran BibTeX style support page is at:
% http://www.michaelshell.org/tex/ieeetran/bibtex/
%\bibliographystyle{IEEEtran}
% argument is your BibTeX string definitions and bibliography database(s)
%\bibliography{IEEEabrv,../bib/paper}
%

\bibliographystyle{IEEEtran}
\bibliography{IEEEfull,conffull,../references}

% biography section
% 
% If you have an EPS/PDF photo (graphicx package needed) extra braces are
% needed around the contents of the optional argument to biography to prevent
% the LaTeX parser from getting confused when it sees the complicated
% \includegraphics command within an optional argument. (You could create
% your own custom macro containing the \includegraphics command to make things
% simpler here.)
%\begin{IEEEbiography}[{\includegraphics[width=1in,height=1.25in,clip,keepaspectratio]{mshell}}]{Michael Shell}
% or if you just want to reserve a space for a photo:

% \begin{IEEEbiography}{Michael Shell}
% Biography text here.
% \end{IEEEbiography}

% if you will not have a photo at all:
% \begin{IEEEbiographynophoto}{John Doe}
% Biography text here.
% \end{IEEEbiographynophoto}

% insert where needed to balance the two columns on the last page with
% biographies
%\newpage

% \begin{IEEEbiographynophoto}{Jane Doe}
% Biography text here.
% \end{IEEEbiographynophoto}

% You can push biographies down or up by placing
% a \vfill before or after them. The appropriate
% use of \vfill depends on what kind of text is
% on the last page and whether or not the columns
% are being equalized.

%\vfill

% Can be used to pull up biographies so that the bottom of the last one
% is flush with the other column.
%\enlargethispage{-5in}

% that's all folks
\end{document}